\documentclass[useAMS,usenatbib]{mn2e}

\usepackage{graphicx}
\usepackage{longtable}
\newcommand{\APer}{$\alpha$ Per}

\title[The Alpha Per astrometric and photometric mass function]
{Astrometric and photometric initial mass functions from the UKIDSS Galactic 
Clusters Survey: II The Alpha Persei open cluster
\thanks{Based on observations made with the United Kingdom Infrared
Telescope, operated by the Joint Astronomy Centre on behalf of the
U.K. Particle Physics and Astronomy Research Council.}}
\author[N. Lodieu et al.]{N. Lodieu$^{1,2}$\thanks{E-mail: nlodieu@iac.es}, 
N. R. Deacon$^{3}$, N. C. Hambly$^{4}$, S. Boudreault$^{1,2}$ \\
$^{1}$Instituto de Astrof\'isica de Canarias (IAC), V\'ia L\'actea s/n,
E-38205 La Laguna, Tenerife, Spain \\
$^{2}$ Departamento de Astrof\'isica, Universidad de La Laguna (ULL),
E-38205 La Laguna, Tenerife, Spain \\
$^{3}$Max-Planck-Institut f\"ur Astronomie, K\"onigstuhl 17, 69117, 
Heidelberg, Germany \\
$^{4}$Scottish Universities' Physics Alliance (SUPA),
Institute for Astronomy, School of Physics and Astronomy, University of Edinburgh,
\\Royal Observatory, Blackford Hill, Edinburgh EH9 3HJ, UK} 

\begin{document}

\date{Accepted \today. Received \today; in original form \today}

\pagerange{\pageref{firstpage}--\pageref{lastpage}} \pubyear{2005}

\maketitle

\label{firstpage}

%
%
\begin{abstract}
We present the results of a deep ($J$ = 19.1 mag) infrared ($ZYJHK$) survey over
the full \APer{} open cluster extracted from the Data Release 9 of the UKIRT 
Infrared Deep Sky Survey Galactic Clusters Survey. We have selected
$\sim700$ cluster member candidates in $\sim$56 square degrees in \APer{} 
by combining photometry in five near-infrared passbands and proper motions 
derived from the multiple epochs provided by the UKIDSS GCS DR9\@.
We also provide revised membership for all previously published \APer{} low-mass 
stars and brown dwarfs recovered in GCS based on the new photometry and 
astrometry provided by DR9\@. We find no evidence of $K$-band variability in 
members of \APer{} with dispersion less than 0.06--0.09 mag. 
We employed two independent but complementary methods to derive the cluster 
luminosity and mass functions: a probabilistic analysis and a more standard 
approach consisting of stricter astrometric and photometric cuts. We find that 
the resulting luminosity and mass functions obtained from both methods are 
consistent. We find that the shape of the \APer{} mass function is similar to 
that of the Pleiades although the characteristic mass may be higher after
including higher mass data from earlier studies (the dispersion is comparable).
We conclude that the mass functions of \APer{}, the Pleiades, and Praesepe
are best reproduced by a log-normal representation similar to the system 
field mass function although with some variation in the characteristic mass
and dispersion values.
\end{abstract}


%
%
\section{Introduction}
\label{APer:intro}

The shape of the Initial Mass function (IMF) is of prime importance to 
understand the processes 
responsible for the formation of stars and brown dwarfs. The definition 
and the first estimate of the IMF was presented in \citet{salpeter55}. 
Our knowledge of the IMF has now improved both at the high-mass and 
low-mass ends. The mass spectrum in open clusters and in the field, defined 
as dN/dM$\propto$M$^{-\alpha}$ ($\alpha$ is the exponent of the power law and
equivalent to $x+1$, where $x$ is the slope of the logarithmic mass function), is currently best fit by a three segment power law 
with $\alpha$ = 2.7 for stars more massive than 1 M$_{\odot}$, $\alpha$ = 2.2
between 1 and 0.5 M$_{\odot}$, and $\alpha$ = 1.3$\pm$0.5 in the 
0.5--0.08 M$_{\odot}$ mass range \citep{kroupa02}. Alternatively, a log--normal
function with a characteristic mass around 0.2--0.25 M$_{\odot}$ and dispersion 
$\sim0.55$ \citep{chabrier03,chabrier05a} provides a good match to current 
observations for the system mass function in the field. The advent of 
large-scale optical and near-infrared surveys towards open clusters extended 
the mass spectrum  to the substellar regime but a consensus has yet to 
emerge on the detailed shape.

\APer{} is one of the few open star clusters within 200 pc of the Sun 
and younger than 200 Myr. The cluster is located to the north-east of 
the F5V supergiant Alpha Persei at a distance of $\sim$175--190 pc 
\citep{pinsonneault98,robichon99} with a revised distance of 
172.4$\pm$2.7 pc from the re-reduction of the Hipparcos data 
\citep{vanLeeuwen09}. The cluster members have solar 
metallicity \citep{boesgaard90} and the extinction along the line of 
sight is estimated as A$_{V}$ = 0.30 mag with a possible differential 
extinction \citep{prosser92}. It has been well studied, though less 
frequently than the Pleiades due to a smaller proper motion 
\citep[($\mu_{\alpha}\cos{\delta}$,$\mu_{\delta}$) = ($+$22.73,$-$26.51) mas/yr; ][]{vanLeeuwen09} 
and a much lower galactic latitude (b = $-$7$^{\circ}$ vs.\ $-$24$^{\circ}$).
Despite being further away than the Pleiades (170 pc vs.\ 120 pc), \APer{} is 
a good target for substellar studies because it is younger than the Pleiades 
(85$\pm$10 Myr vs.\ 125$\pm$8 Myr), placing the lithium depletion boundary at 
$I_{c} \sim$ 17.7--17.8 mag for both clusters.

Multi-wavelength surveys and spectroscopic follow-up observations have been 
performed in \APer{} to extract a clean sequence of cluster members from 
high-mass stars down to brown dwarfs. The first proper motion survey in the 
cluster was performed by \citet{heckmann56} and complemented by photometry 
from \citet{mitchell60}, yielding about 60 probable members (HE objects) 
whose final membership was revised by \citet{prosser92}. The membership of 
additional candidates proposed by \citet{fresneau80} was subsequently 
established by \citet{prosser92}. Lower mass members (AP sources) were 
extracted by \citet{stauffer85} and \citet{stauffer89b} on the basis of 
their proper motion, photometry, and spectral characteristics.
\citet{prosser92} examined the Palomar photographic plates to extract new 
low-mass proper motion and photometric members down to a spectral type of 
M4 over a 6$^{\circ}$ by 6$^{\circ}$ field. Additional low-mass photometric 
candidates were reported from a deeper optical survey in a smaller area 
\citep{prosser94} as well as from X-rays observations with ROSAT
\citep{randich96,prosser96b,prosser98a,prosser98b}.
The first brown dwarf candidates were spectroscopically confirmed by 
\citet{stauffer99}, yielding a lithium age of 90$\pm$10 Myr, twice 
the turn-off main-sequence age \citep[50 Myr;][]{mermilliod81}.
A revised value of the age derived from the lithium method was published 
by \citet{barrado04a}, estimated to 85$\pm$10 Myr. A deep optical survey 
complemented by near-infrared photometry extended the cluster sequence down 
to 0.03 M$_{\odot}$ \citep{barrado02a}.
The best fit of the slope of the mass function was obtained for a power 
law index $\alpha$ = 0.59$\pm$0.05 over the 0.3--0.035 M$_{\odot}$ mass range, 
in agreement with estimates in the Pleiades \citep{dobbie02a,moraux03} at 
that time. A wider survey based on photographic plates by \citet{deacon04}
derived a power law index $\alpha$ of 0.86 (0.67--1.00) over the 
1.0--0.2 M$_{\odot}$ range from a sample of high probability members over 
$\sim$250 square degrees. Finally, \citet{lodieu05a} extracted about 20 new 
infrared photometric candidates from a deep $K$-band survey of 0.7 square 
degree previously covered in the optical by \cite{barrado02a}. Additionally, 
24 probable candidates from \citet{barrado02a} were confirmed as 
spectroscopic members with masses between 0.4 and 0.12 M$_{\odot}$.

The UKIRT Infrared Deep Sky Survey \citep[UKIDSS;][]{lawrence07} is a deep 
large-scale infrared survey conducted with the wide-field camera WFCAM 
\citep{casali07} on UKIRT (Mauna Kea, Hawai'i). The survey is subdivided 
into 5 components: the Large Area Survey, the Galactic Clusters Survey 
(hereafter GCS), the Galactic Plane Survey, the Deep Extragalactic Survey, 
and the Ultra-Deep Survey. The GCS aims at covering $\sim$1000 square degrees 
in 10 star-forming regions and open clusters down to $K$ = 18.4 mag at two 
epochs. The main scientific driver of the survey is to study the IMF and its 
dependence with environment in the substellar regime 
using an homogeneous set of low-mass stars and brown dwarfs over a large area 
in several regions.

In this paper we present the \APer{} mass function over $\sim$56 square 
degrees derived from the UKIDSS GCS Data Release 9 (DR9). This is the 
second paper of its kind after the analysis of the Pleiades cluster presented 
in \citet{lodieu12a}. In Section \ref{APer:ukidss_GCSDR9} we present the 
photometric and astrometric dataset employed to extract member candidates in 
\APer{}. In Section \ref{APer:status_old_cand} we review the list of previously 
published members recovered by the UKIDSS GCS DR9 and revise their membership.
In Section \ref{APer:new_cand}  we outline two methods for deriving the cluster 
luminosity function. One method relies on a relatively conservative photometric 
selection followed by the calculation of formal membership probabilities based 
on object positions in the proper motion vector point diagram 
(Section \ref{APer:new_cand_probabilistic}). The second method applies a more 
stringent colour cut followed by an astrometric selection based on the formal 
errors on the proper motions for each photometric candidate compared 
to that of the cluster (Section \ref{APer:new_cand_phot_PM}) for which we
test the level of contamination (Sect.\ \ref{APer:contamination}).
In Section \ref{APer:variability} we discuss the $K$-band variability
of cluster member candidates in \APer{}.
In Section \ref{APer:IMF} we derive the cluster luminosity and (system) mass 
function and compare it to other clusters studied as part of the GCS
(Pleiades and Praesepe), and the field.

%
%
\section{The UKIDSS GCS in \APer{}}
\label{APer:ukidss_GCSDR9}

The UKIDSS GCS DR9 released $\sim$56 square degrees observed in five 
passbands \citep[$ZYJHK$;][]{hewett06} in the \APer{} open cluster over a 
region defined by RA=44--60 degrees and dec=44--54 degrees. 

We have selected all good quality point sources in \APer{} detected in at 
least {$JHK$1 (where $K$1 stands for the first $K$-band epoch) and, where 
available, in $Z$, $Y$, and $K$2 (second $K$-band epoch).
We imposed a request on point sources only in $JHK$ and pushed the
completeness towards the faint end by imposing limits on the {\tt{ClassStat}}
parameters (between $-$3 and $+$3) which classify the point--likeness of an image.
The Structured Query Language (SQL) query used to select sources along the
line of sight of the \APer{} is identical to the query used for the Pleiades
\citep{lodieu12a}.  The SQL query includes the cross-matches with 2MASS 
\citep{cutri03,skrutskie06} to compute proper motions for all sources brighter 
than the 2MASS 5$\sigma$ completeness limit at $J$ = 15.8 mag as well as
the selection of proper motions from multiple epochs provided by the GCS\@.
We used the GCS proper motion measurements in this work as they 
are more accurate due to the homogeneous coverage, completeness, and spatial 
resolution of the UKIDSS images and the detailed relative astrometric mapping
employed \citep{collins12}, and of course the GCS proper motions are available 
for objects that are too faint for 2MASS. We limited our selection to sources 
fainter than $Z$ = 11.6, $Y$ = 11.4, $J$ = 11.0, $H$ = 11.5, $K$1 = 10.0, 
$K$2 = 10.4 mag to avoid saturated point sources.
The completeness limits, taken as the magnitude where the straight line
fitting the shape of the number of sources as a function of magnitude falls
off, are $Z$ = 20.0, $Y$ = 19.6, $J$ = 19.1, $H$ = 18.4, $K$1 = 17.6, and 
$K$2 = 18.1 mag (Fig.~\ref{fig_APer:compl_DR9}).

The query returned 2,643,045 sources with $J$ = 11.0--21.2 mag over $\sim$56
square degrees towards the \APer{} cluster. The full coverage is displayed in 
Fig.~\ref{fig_APer:GCScoverage} and the resulting ($Z-J$,$Z$) colour-magnitude
diagram is shown in Fig.~\ref{fig_APer:ZJZcmd_alone_DR9} along with previously 
published member candidates (black filled dots). Note that theoretical 
isochrones plotted in this paper were specifically computed for the WFCAM set 
of filters at an age of 90 Myr (downloaded from France Allard's 
webpage)\footnote{France Allard's Phoenix web simulator can be found at 
http://phoenix.ens-lyon.fr/simulator/index.faces}. We combined the NextGen
and DUSTY isochrones for effective temperatures above and below 2700\,K,
respectively, to convert magnitudes into masses (Section~\ref{APer:IMF}).

%
%
\begin{figure}
   \centering
   \includegraphics[width=\linewidth]{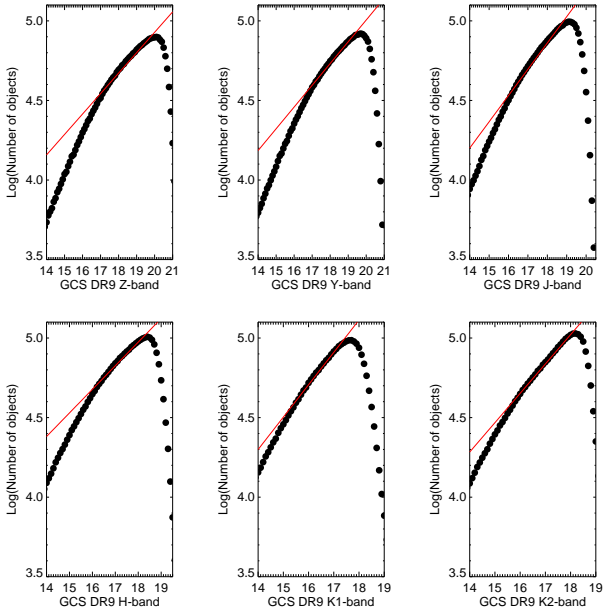}
   \caption{Completeness of the GCS DR9 dataset in the \APer{} cluster in 
each of the six filters. The polynomial fit of order 2 is shown as a red line and 
defines the 100\% completeness limit of the GCS DR9 in each passband.}
   \label{fig_APer:compl_DR9}
\end{figure}
%

%
%
\begin{figure}
   \includegraphics[width=\linewidth]{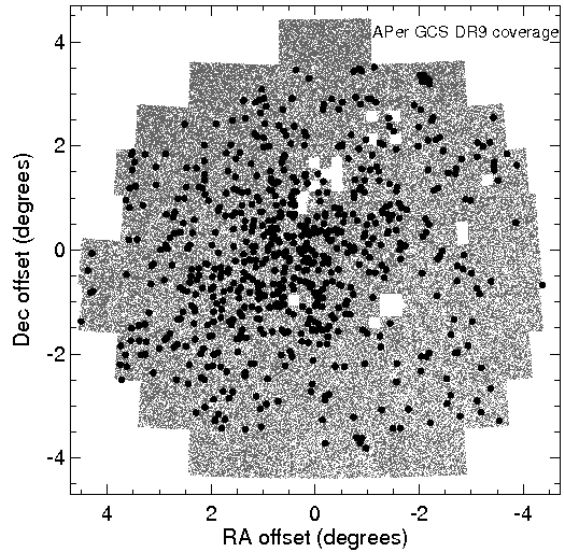}
   \caption{
Coverage from the UKIDSS GCS DR9 in the \APer{} open cluster in the standard
angular plane coordinates $(\xi,\eta)$ choosing (ra,dec) = (51, 49) degrees as the
cluster centre. The total area covered is about 56 square degrees. 
The holes present in the coverage are due to the rejection of some tiles 
after quality control. GCS DR9 member candidates identified in this work
are overplotted as black filled dots.
}
   \label{fig_APer:GCScoverage}
\end{figure}
%

%
%
\begin{figure}
   \includegraphics[width=1.00\linewidth]{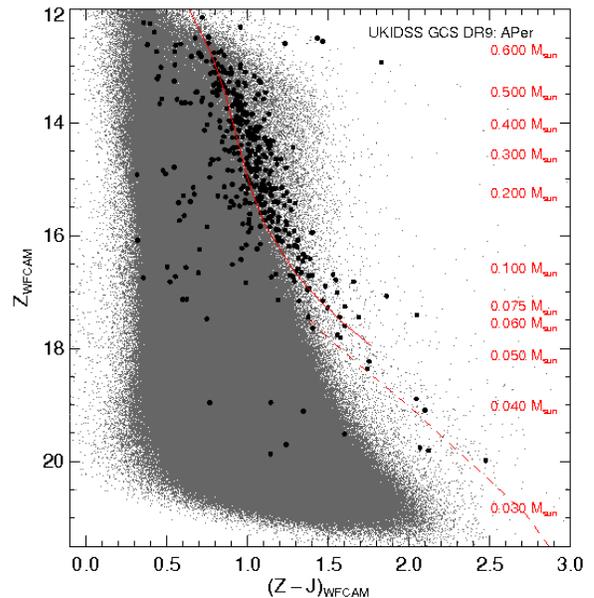}
   \caption{($Z-J$,$Z$) CMD for $\sim$56 square degrees in the \APer{} extracted
from the UKIDSS Galactic Cluster Survey Data Release 9\@. Previously published
member candidates in \APer{} are overplotted as filled dots. The mass scale is 
shown on the right hand side of the diagrams and extends down to 
0.03 M$_{\odot}$, according to the NextGen and DUSTY models assuming an age 
of 90 Myr and a distance of 172.4 pc \citep{baraffe98,chabrier00c}.
}
   \label{fig_APer:ZJZcmd_alone_DR9}
\end{figure}

%
%
\section{Cross-match with previous surveys}
\label{APer:status_old_cand}

There are 455 probable members known in \APer{} extracted from previous 
proper motion and optical surveys
\citep{heckmann56,mitchell60,fresneau80,stauffer85,stauffer89b,prosser92,prosser94,prosser98a,prosser98b,stauffer99,barrado02a,lodieu05a},
and an additional 300 high-probability (p $\geq$ 60\%) member
candidates from \citet{deacon04}.

%
%
\begin{table*}
  \caption{Updated membership of member candidates identified in \APer{}
by earlier studies and recovered in the GCS DR9 sample. Papers {\em studying}
\APer{} over the past decades and considered in this work are:
\citep{heckmann56,mitchell60,fresneau80,stauffer85,stauffer89b,prosser92,prosser94,prosser98a,prosser98b,stauffer99,barrado02a,deacon04,lodieu05a}.
Columns 2 and 3 give the numbers of sources published by the reference
given in Column 1 and the numbers of sources recovered in GCS DR9, respectively.
Column 4 (named No\_DR9)  is subdivided into several columns to give the 
reasons why some of the sources from earlier studies are not covered:
``Bright'' stands for objects brighter than the GCS saturation limits,
``Outside'' stands for sources outside the GCS DR9 coverage,
``No\_mag'' stands for sources missing at least one of the $J$, $H$, or $K$ magnitudes,
``$>$3$''$'' stands for sources beyond the 3 arcsec matching radius used in our study,
and ``Flag'' stands for sources whose GCS flags are too bad to be included in our catalogue of point sources.
Columns 5 and 6 give the numbers of high-probability members (p$\geq$40\%) and
non members (NM) according to our probabilistic approach (first number) and
method \#2 (second number). The last column gives the percentages of sources 
recovered in the GCS DR9 (ratio DR9/All). 
}
  \label{tab_APer:early_summary}
  \begin{tabular}{@{\hspace{0mm}}l @{\hspace{3mm}}c @{\hspace{3mm}}c @{\hspace{2mm}}c @{\hspace{2mm}}c @{\hspace{2mm}}c @{\hspace{2mm}}c @{\hspace{2mm}}c @{\hspace{3mm}}c @{\hspace{3mm}}c @{\hspace{3mm}}c@{\hspace{0mm}}}
  \hline
Survey        &  All  & DR9 & \multicolumn{5}{|c|}{No\_DR9} & Memb  &  NM  & \%   \cr
              &       &     & Bright & Outside & No\_mag & $>$3$''$ & Flag &       &      &      \cr
  \hline
Heckmann1956      &  144  (78) &   7 & 65 &  1 & 71 &  0 &  0 &  0/0     &   0/7    &    4.9  (9.0) \cr
Fresneau1980      &   56  (26) &   2 & 28 &  0 & 26 &  0 &  0 &  0/0     &   0/2    &    3.6 (46.4) \cr
Prosser1992       &  148  (96) &  44 & 34 & 18 & 24 & 25 &  1 & 28/31    &  16/13   &   29.7 (45.8) \cr
Prosser1994       &   31  (30) &  23 &  0 &  1 &  2 &  3 &  2 & 12/14    &  11/9    &   74.2 (76.7) \cr
Prosser1998a      &   89  (62) &  43 & 27 &  0 & 12 &  2 &  5 & 15/11    &  28/32   &   48.3 (69.4) \cr
Prosser1998b      &   70  (41) &  28 & 28 &  2 & 11 &  2 &  0 & 10/15    &  18/13   &   40.0 (68.3) \cr
Stauffer1999      &   28  (28) &  23 &  0 &  0 &  0 &  0 &  0 &  9/10    &  14/13   &   82.1 (82.1) \cr
Barrado2002\_prob &   56  (56) &  48 &  0 &  0 &  6 &  6 &  0 & 25/32    &  23/16   &   85.7 (85.7) \cr
Barrado2002\_poss &   13  (13) &   7 &  0 &  1 &  1 &  3 &  1 &  4/4     &   3/3    &   53.8 (53.8) \cr
Barrado2002\_NM   &   29  (29) &  15 &  0 &  1 &  2 &  3 &  0 &  3/3     &  12/12   &   51.7 (51.7) \cr
Deacon2004        &  302 (258) & 244 & 24 & 20 &  8 &  0 &  6 &154/149   &  90/95   &   80.8 (94.6) \cr
Lodieu2005        &   39  (18) &   5 &  0 & 16 &  8 &  9 &  1 &  2/4     &   3/1    &   12.8 (27.8) \cr
 \hline
\end{tabular}
\end{table*}

We cross-matched catalogues from earlier studies with our full sample
of over $\sim$2.5 million sources retrieved from GCS DR9 to locate the
cluster sequence in various colour-magnitude diagrams. We recovered a 
total of 426 known members in \APer{} after removing multiple detections 
present in various catalogues (Table \ref{tab_APer:early_DR9}).
The numbers and percentages in brackets in the second and sixth column of 
Table \ref{tab_APer:early_summary} consider previously published sources
lying in the magnitude range probed by the GCS\@. We also made a detailed
analysis of the 629 previously known members not recovered by our SQL query.
The numbers are given in the fourth column of Table \ref{tab_APer:early_DR9}
which is divided into five sub-columns. Most of these sources are either
missing an image in $J$, $H$, or $K$1 or are not covered by the GCS (223 or 
35.5\%) or are brighter than the saturation limits set in our query (205 or 
32.6\%) or are very likely proper motion non members (48 or 7.6\%).

%
%
%
\section{New substellar members in \APer{}}
\label{APer:new_cand}
%
%
\subsection{Probabilistic approach}
\label{APer:new_cand_probabilistic}
%

%
%
\subsubsection{Method}
\label{APer:new_cand_probabilistic_method}

In this section we outline the probabilistic approach we employed to select 
low-mass stars and brown dwarf member candidates in \APer{} using photometry 
and astrometry from the UKIDSS GCS DR9\@. This method is described in detail
in \citet{deacon04} and \citet{lodieu07c}. The main steps are:
\begin{enumerate}
\item Define the cluster sequence using candidates published in the literature
within the area covered by the GCS DR9
\item Make a conservative cut in the ($Z-J$,$Z$) diagram to include known
members and any new cluster member candidates defined as
($Z$\,$\geq$\,16.5 and $Z$\,$\leq$(11.5$+$\,5.0$\times$\,($Z-J$)) OR ($Z$\,$\leq$\,16.5 and $Z$\,$\leq$(8.5\,$+$\,8.0$\times$\,($Z-J$))
displayed as solid black lines on the top-left panel in Fig.~\ref{fig_APer:YJK_cmds}.
\item Analyse the vector point diagram in a probabilistic manner to assign 
a membership probability for each photometric candidate with a proper motion 
measurement (Section \ref{APer:new_cand_probabilistic_proba}).
\item Derive the luminosity and mass function by summation of membership
probabilities to provide a statistically complete sample.
\end{enumerate}

%
%
\begin{figure}
   \includegraphics[width=\linewidth]{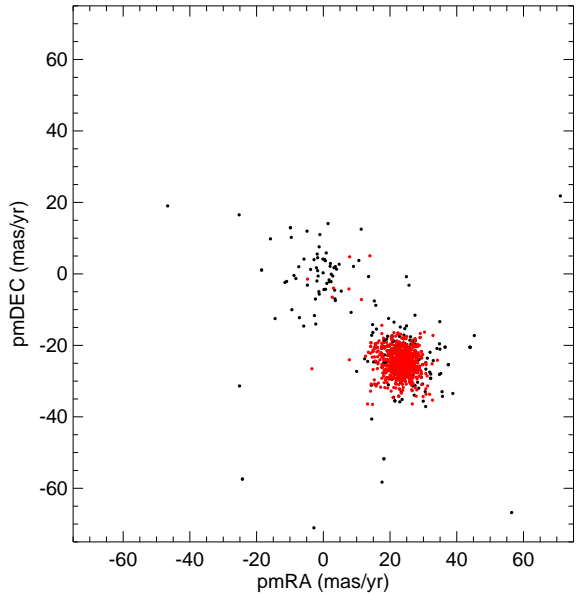}
   \caption{Vector point diagram showing the proper motion in right ascension
(x-axis) and declination (y-axis) for previously known member candidates
recovered by the GCS DR9 (black dots) and the new member candidates
selected with method \#2 (red dots).
}
   \label{fig_APer:diagram_VPD}
\end{figure}
%

%
%
\subsubsection{Membership probabilities}
\label{APer:new_cand_probabilistic_proba}

In order to calculate formal membership probabilities we have used the same 
technique as in \citet{lodieu12a} to fit distribution functions to proper 
motion vector point diagrams \citep{hambly95}. The technique differs slightly
from the original method presented in \citet{deacon04} and \citet{lodieu07c}
in that the value of sigma ($\sigma$) is fixed by the formal astrometric errors
propagated from the centroiding errors given by the source extraction software
employed upstream of the WFCAM Science Archive \citep{hambly08}, the main repository
of UKIDSS data.

We refer the reader to the above cited papers}for more details and additional equations.
First, we rotated the vector point diagram so the cluster lies on the 
y-axis, assuming a proper motion of (22.73,$-$26.51) mas/yr for \APer{} 
\citep{vanLeeuwen09} after applying a very conservative photometric selection
in the ($Z-J$,$Z$) colour-magnitude diagram. We also note that we used the
following rotation for the vector point diagram: 
\begin{itemize}
\item $\mu_{x'}$\,=\,$\cos$(0.77$\times$PI)\,$\times$\,$\mu_{x}$\,$-$\,$\sin$(0.77$\times$PI)\,$\times$\,$\mu_{y}$
\item $\mu_{'y}$\,=\,$\cos$(0.77$\times$PI)\,$\times$\,$\mu_{x}$\,$+$\,$\sin$(0.77$\times$PI)\,$\times$\,$\mu_{y}$
\end{itemize}

We have assumed that there are two contributions to the total distribution 
$\phi(\mu_{x},\mu_{y})$, one from the cluster, $\phi_{c}(\mu_{x},\mu_{y})$, 
and one from the field stars, $\phi_{f}(\mu_{x},\mu_{y})$. The fitting region 
was delineated by $-$50 $< \mu_{x} <$ 50 mas/yr and $-$50 $< \mu_{y} <$ 50 
mas/yr. These were added by means of a field star fraction $f$.

We characterised the cluster distribution as a 
bivariate gaussian with a single standard deviation $\sigma$ and mean proper 
motion values in each axis $\mu_{xc}$ and $\mu_{yc}$.
The field star distribution was fitted by a univariate gaussian 
in the $x$ axis (with standard deviation $\Sigma_{x}$ and mean $\mu_{xf}$) and 
a declining exponential in the $y$ axis with a scale length $\tau$.
The use of a declining exponential is a standard method \citep[e.g.][]{jones91} 
and is justified in that the field star distribution is not simply a 
circularly-symmetric error distribution (i.e.\ capable of being modelled as a 2d 
Gaussian) - rather there is a prefered direction of real field star motions 
resulting in a characteristic velocity ellipsoidal signature, i.e.\ a 
non-Gaussian tail, in the vector point diagram. This is best modelled (away 
from the central error-dominated distribution) as an exponential in the 
direction of the antapex (of the solar motion).

The best fitting set of parameters were chosen using a maximum likelihood 
method \citep[see][]{deacon04}. However in a deviation from this method we 
did not fit for the standard deviation of the cluster proper motions ($\sigma$). 
Instead we calculated the mean astrometric error for all objects in each 
magnitude range and used this as our cluster standard deviation.
This fitting process was tested by \citet{deacon04} where
simulated data sets were created and run through the fitting process
to recover the input parameters. These tests produced no significant 
offsets in the parameter values \citep[see Table 3 and Appendix A 
of][for results and more details on the procedure]{deacon04}.
Hence, we calculated the formal membership probabilities as,

\begin{equation}
p=\frac{\phi_{c}}{f\phi_{f}+(1-f)\phi_{c}}
\end{equation}

%
%
\begin{table}
  \caption{Summary of the results after running the programme
to derive membership probabilities. For each $Z$ magnitude range,
we list the number of stars used in the fit (Nb), the field star 
fraction f, and parameters describing the cluster and field star 
distribution. Units are in mas/yr except for the number of 
stars and the field star fraction f. The cluster star distribution
is described by the mean proper motions in the x and y 
directions ($\mu_{x_{c}}$ and $\mu_{y_{c}}$) and a standard
deviation $\sigma$. Similarly, the field star distribution is
characterised by a scale length for the y axis ($\tau$), a
standard deviation $\Sigma_{x}$, and a mean proper motion
in the x direction ($\mu_{x_{f}}$). Note that the value of sigma 
($\sigma$) is fixed by the formal astrometric errors.
}
  \label{tab_APer:prob_results}
  \begin{tabular}{@{\hspace{0mm}}c @{\hspace{3mm}}c @{\hspace{3mm}}c @{\hspace{3mm}}c @{\hspace{3mm}}c @{\hspace{3mm}}c @{\hspace{3mm}}c @{\hspace{3mm}}c @{\hspace{3mm}}c@{\hspace{0mm}}}
  \hline
$Z$ & Nb & f & $\sigma$ & $\mu_{x_{c}}$ & $\mu_{y_{c}}$ & $\tau$ & $\Sigma_{x}$ & $\mu_{x_{f}}$ \\
 \hline
12--13 & 206 & 0.84 & 2.84 & $-$1.64 & 33.24 & 16.56 & 21.67 &  4.76 \\ 
13--14 & 488 & 0.75 & 2.82 & $-$1.98 & 33.91 & 21.32 & 16.27 &  0.78 \\ 
14--15 & 720 & 0.77 & 2.78 & $-$1.73 & 33.99 & 16.83 & 16.21 &  0.60 \\ 
15--16 & 913 & 0.83 & 2.85 & $-$1.74 & 33.47 & 14.69 & 15.05 & -0.50 \\ 
16--17 & 877 & 0.86 & 2.88 & $-$2.15 & 34.30 & 14.68 & 14.66 &  0.21 \\ 
17--18 & 503 & 0.92 & 3.05 & $-$1.42 & 33.35 & 13.71 & 14.27 &  0.08 \\ 
18--19 & 224 & 0.89 & 3.52 & $-$2.39 & 31.24 & 17.35 & 15.38 &  0.98 \\ 
19--20 & 203 & 0.90 & 5.12 & $-$3.12 & 31.62 & 12.39 & 14.81 & -0.39 \\
 \hline
\end{tabular}
\end{table}

%
%
%
\begin{figure*}
   \centering
   \includegraphics[width=0.49\linewidth]{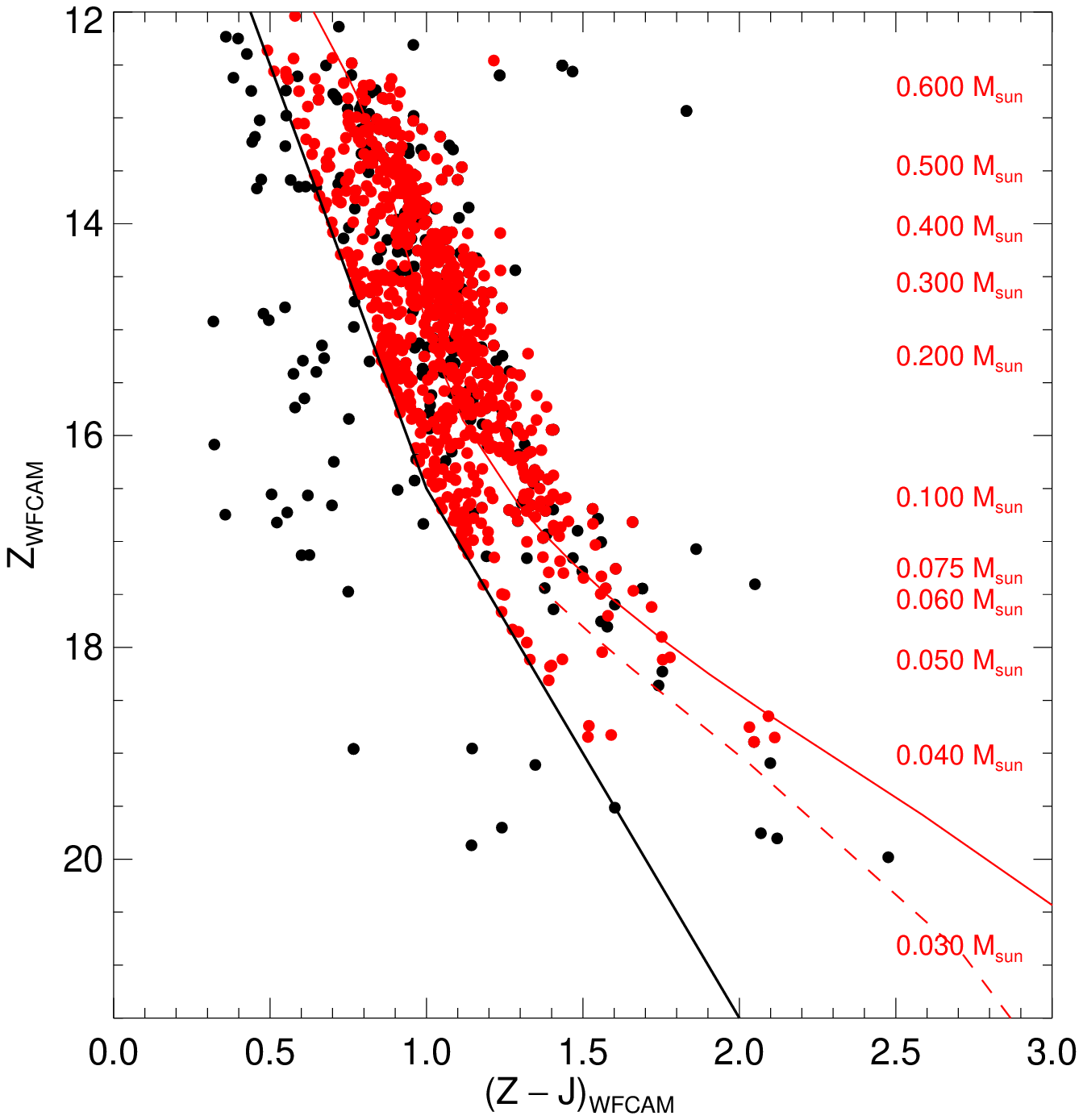}
   \includegraphics[width=0.49\linewidth]{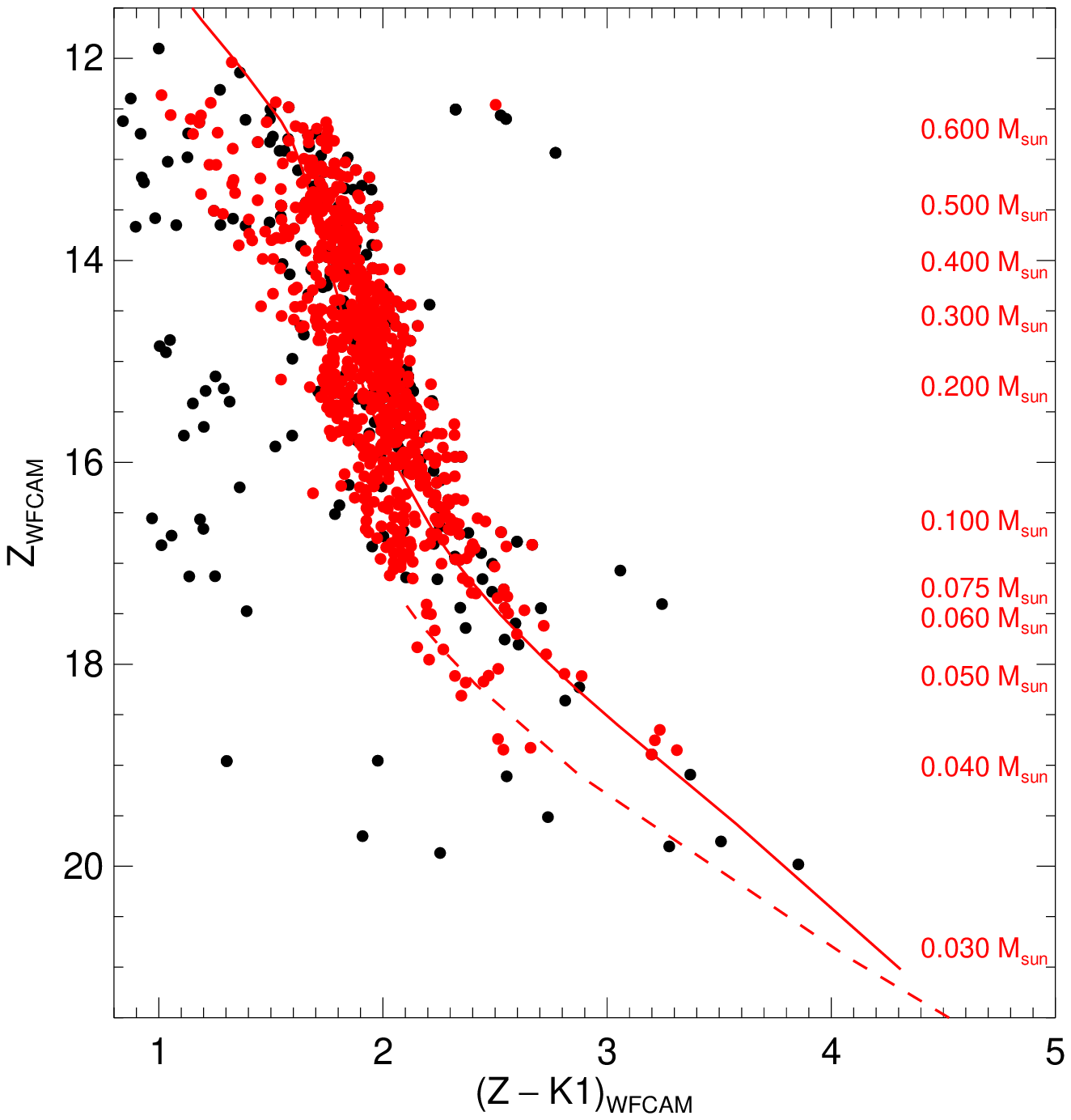}
   \includegraphics[width=0.49\linewidth]{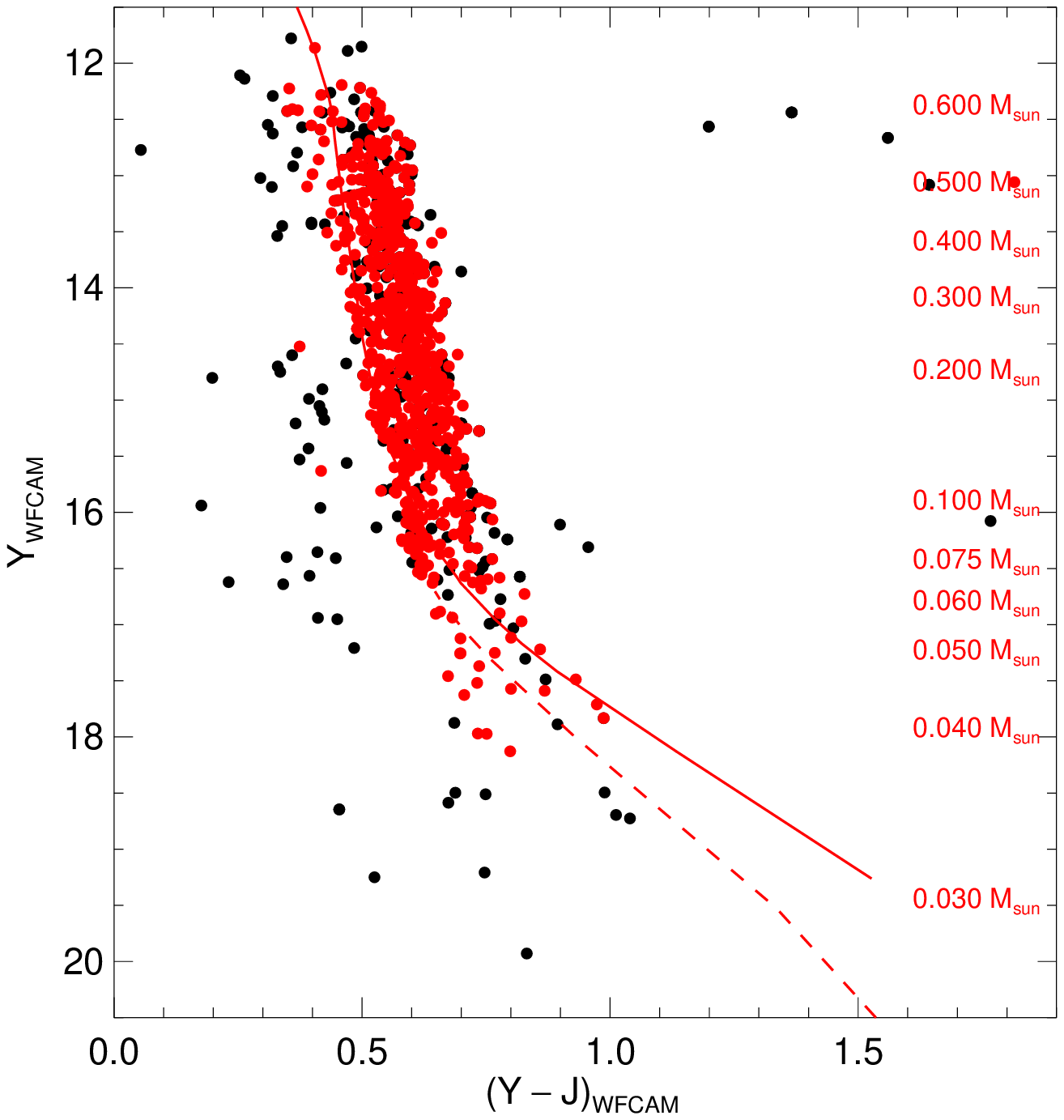}
   \includegraphics[width=0.49\linewidth]{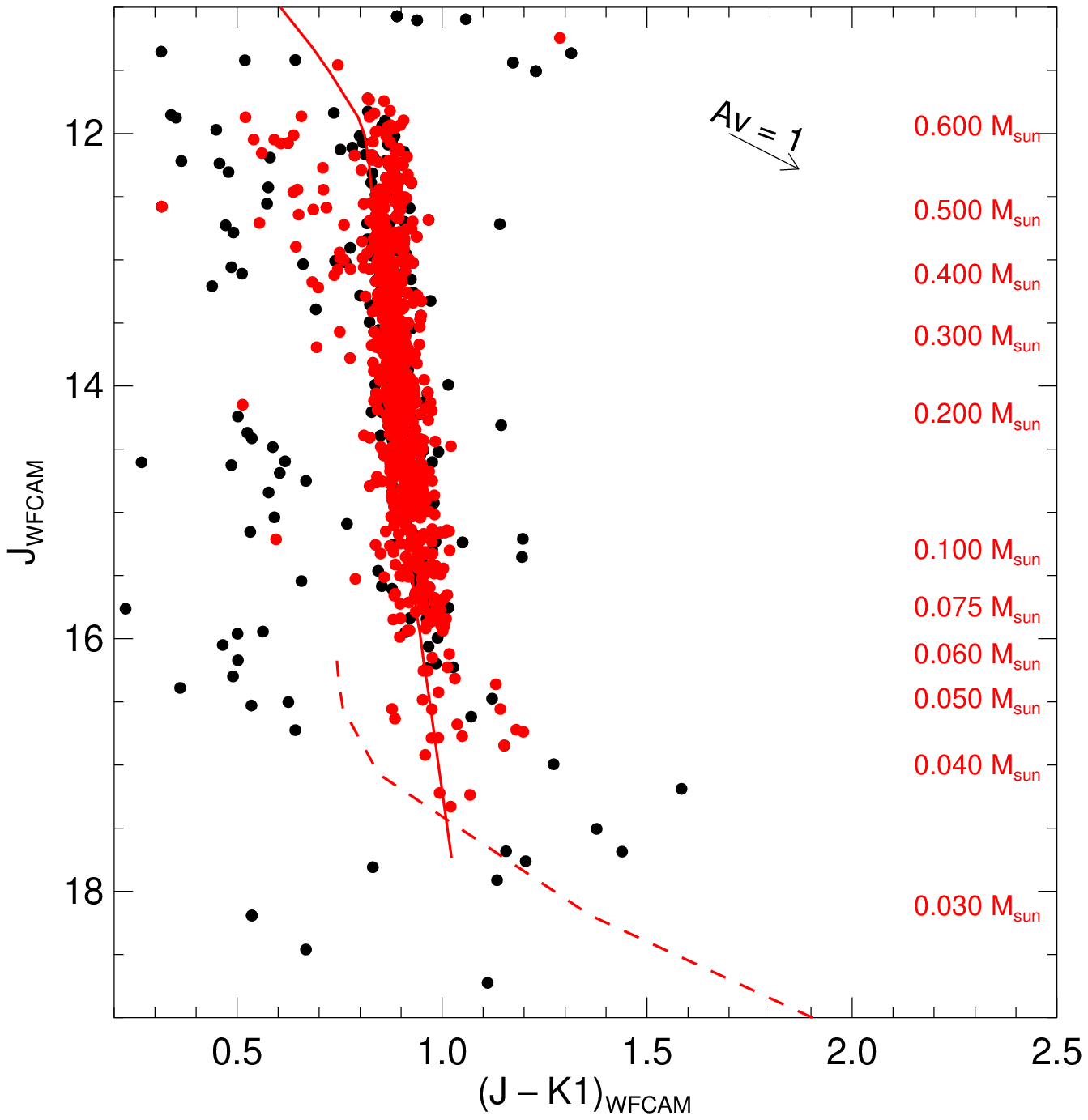}
   \caption{Colour-magnitude diagrams showing the member candidates
   previously reported in \APer{} (black dots) and all candidates extracted
   from our probabilistic analysis, including known ones (red dots).
   {\it{Upper left:}} ($Z-J$,$Z$);
   {\it{Upper right:}} ($Z-K$,$Z$);
   {\it{Lower left:}} ($Y-J$,$Y$);
   {\it{Lower right:}} ($J-K$,$J$).
   Overplotted are the 90 Myr NextGen \citep[solid line;][]{baraffe98} and
   DUSTY \citep[dashed line;][]{chabrier00c} isochrones shifted to a distance
   of 120 pc. The mass scale is shown on the right hand side of the diagrams
   and spans 0.60--0.03 M$_{\odot}$, according to the 90 Myr isochrone models.
   The solid black lines in the upper left diagram represent our conservative 
   photometric cuts used for the probabilistic approach.
}
   \label{fig_APer:YJK_cmds}
\end{figure*}

We split the sample into eight intervals of magnitudes because astrometric
errors are a function of magnitude and also to improve the contrast between
the field stars and the cluster. Each band is one magnitude wide and was fitted
with all seven parameters in the same way as described in \citet{deacon04}. 
There was no fit possible for the 20--21 magnitude bins
because of the small number of sources in this bin. A summary of the 
fitted parameters from the probabilistic analysis described above is given in 
Table \ref{tab_APer:prob_results}.

%
%
\subsubsection{Probabilistic sample}
\label{APer:new_cand_probabilistic_phot}

The probabilistic approach yielded a total sample of 10,176 sources with
membership probabilities assigned to each of them. This sample contains
728 sources with membership probabilities higher than 40\% (including
known ones previously published) listed in
Table \ref{tab_APer:new_highprob}. Tightening this probability threshold to 
50\% and 60\% yields samples of 573 ($\sim$27\% less) and 431 ($\sim$69\% less) 
member candidates in \APer{}, respectively.
These high-probability members are displayed in Fig.~\ref{fig_APer:YJK_cmds}
with previously published candidates in \APer{} plotted in black.

%
%
%
\begin{figure*}
   \centering
   \includegraphics[width=0.49\linewidth]{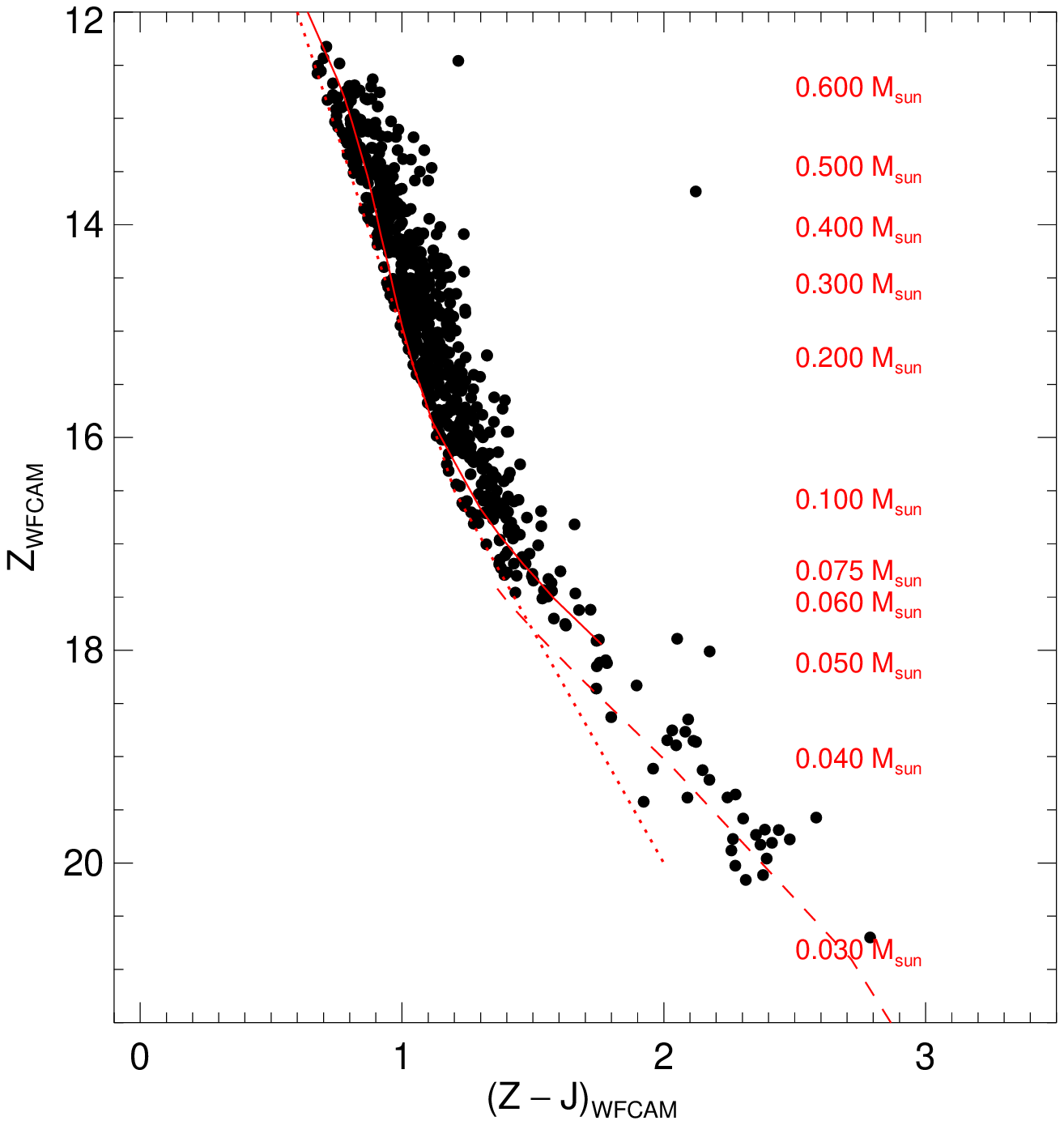}
   \includegraphics[width=0.49\linewidth]{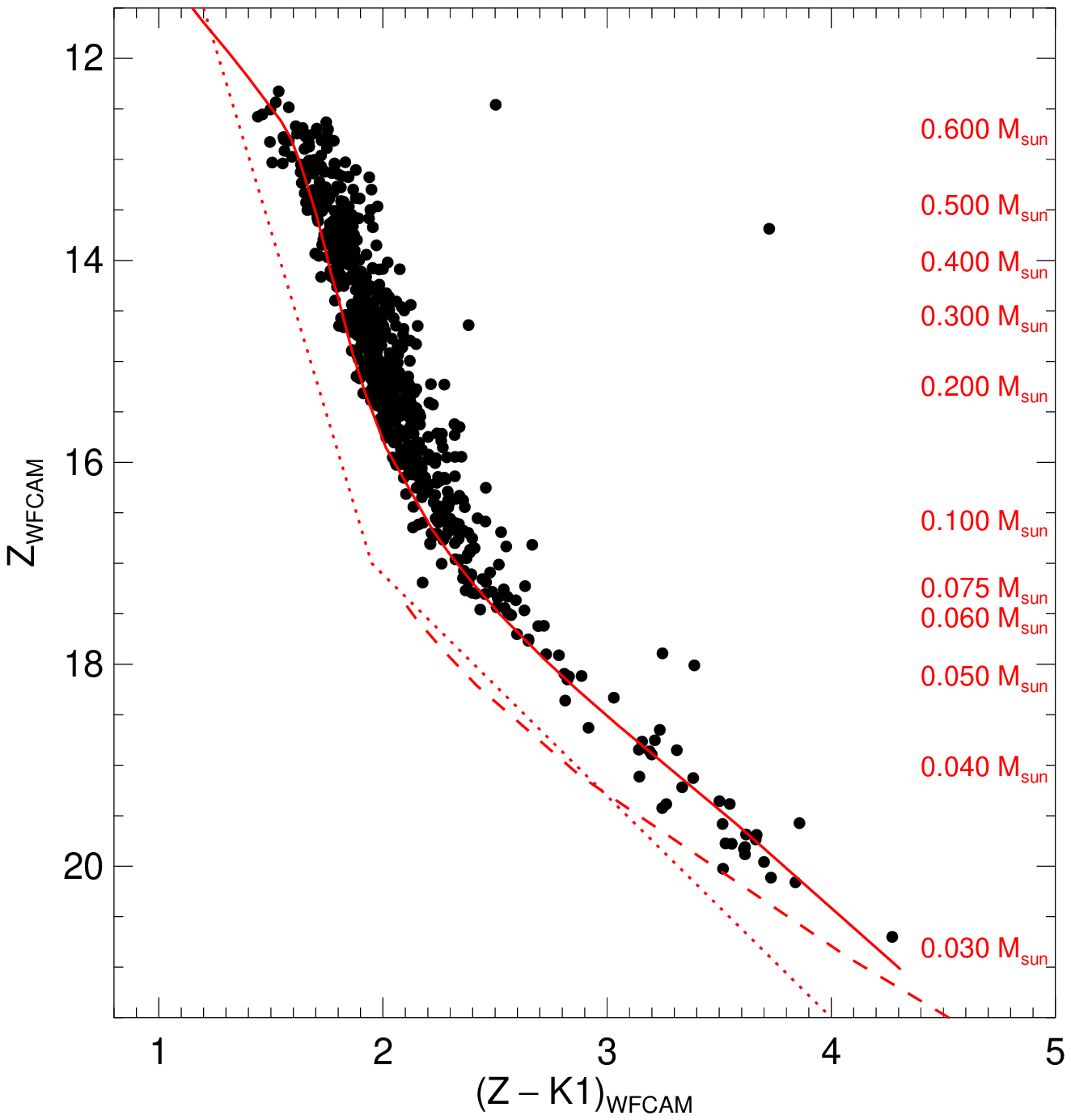}
   \includegraphics[width=0.49\linewidth]{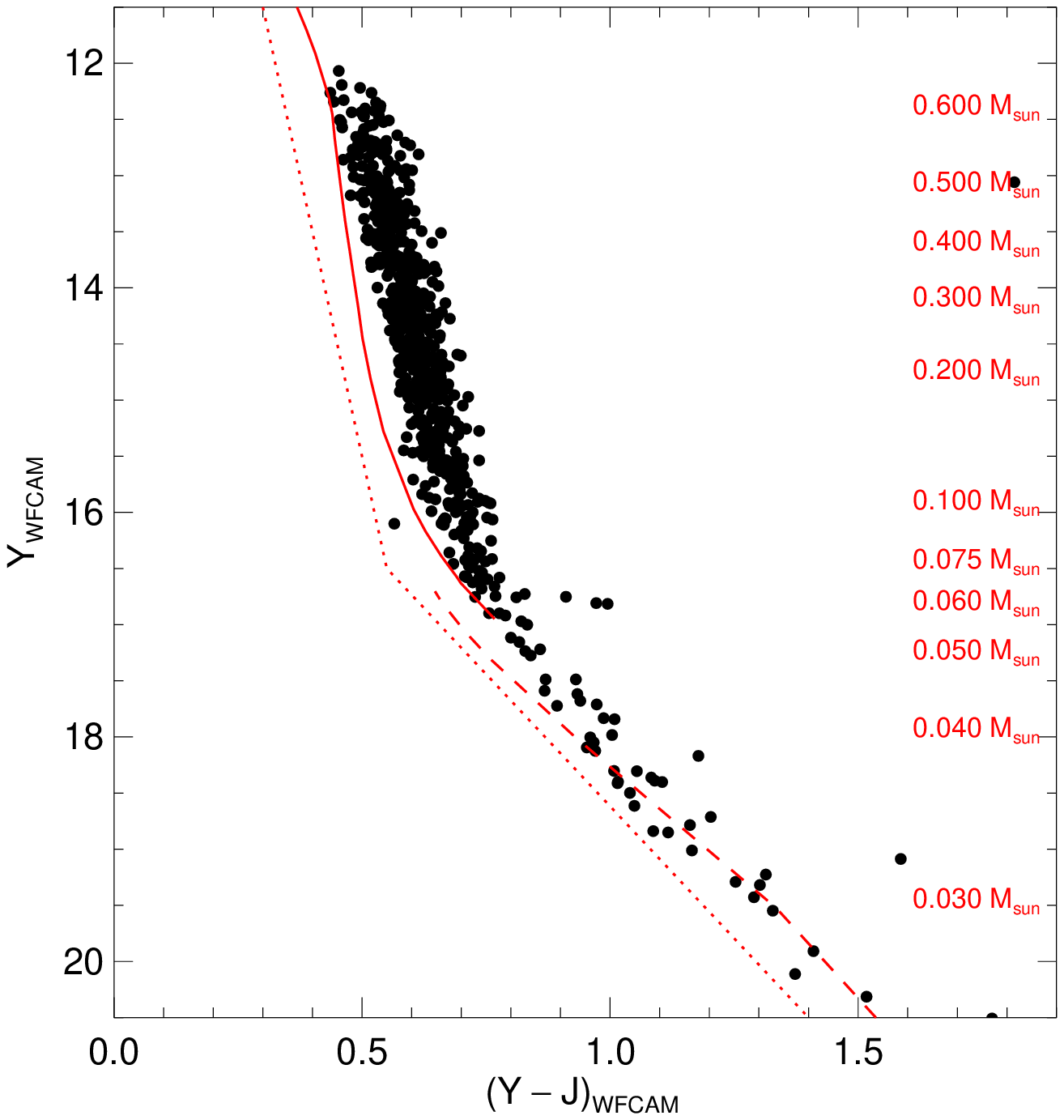}
   \includegraphics[width=0.49\linewidth]{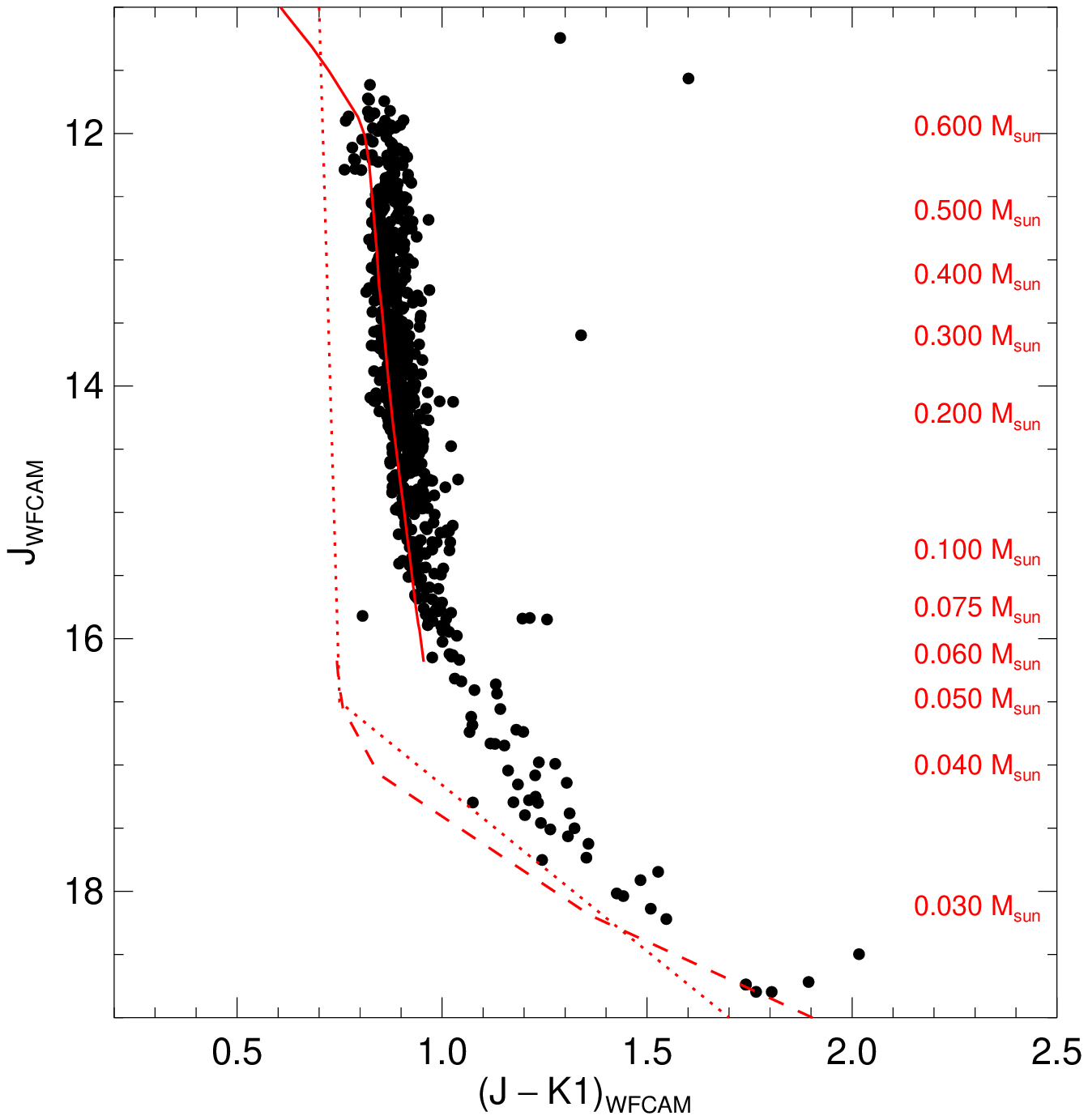}
   \caption{Same as Fig.~\ref{fig_APer:YJK_cmds} but only for member 
candidates in \APer{} selected using method \#2\@. The $YJHK$ and $JHK$-only
detections have been added too for completeness.
}
   \label{fig_APer:YJK_cmds_method2}
\end{figure*}
%

%
\subsection{Photometry and proper motion selection}
\label{APer:new_cand_phot_PM}

In this section we outline a more widely used method (refered to as method \#2 
in the rest of the paper) that we applied to select low-mass and substellar member 
candidates in \APer{}. This procedure consist of selecting cluster candidates by 
applying proper motion selection followed by strict photometric cuts in various 
colour-magnitude diagrams. This alternative method provides and independent test 
of the probabilistic approach presented in the previous section.

The first step was to select all sources with formal errors on the
proper motion within 3$\sigma$ of the mean proper motion of the cluster
(Fig.~\ref{fig_APer:diagram_VPD}), yielding a completeness better than 99\%
assuming normally distributed errors. The main advantage of this method 
is that it does not rely on a single radius
for the proper motion selection but rather takes into account the increasing
uncertainty on the proper motion measurements between the GCS epochs with
decreasing brightness.

Secondly, we plotted several colour-magnitude diagrams 
(Fig.~\ref{fig_APer:YJK_cmds}) to define a series of lines based on the 
position of known \APer{} members identified in earlier studies and published 
over the past decades (Table \ref{tab_APer:early_summary}). Those lines detailed
below are plotted in Fig.~\ref{fig_APer:YJK_cmds_method2} and improve on the
pure proper motion selection. We note that those criteria are similar
to those used for the Pleiades \citep{lodieu12a} because the younger
age of \APer{} compared to the Pleiades is compensated by its larger distance.
\begin{itemize}
\item ($Z-J$,$Z$) = (0.60,12.0) to (1.20,16.5) 
\item ($Z-J$,$Z$) = (1.20,16.5) to (2.00,20.0) 
\item ($Z-K$,$Z$) = (1.20,11.5) to (1.95,17.0)
\item ($Z-K$,$Z$) = (1.95,17.0) to (4.00,21.5)
\item ($Y-J$,$Y$) = (0.30,11.5) to (0.55,16.5) 
\item ($Y-J$,$Y$) = (0.55,16.0) to (1.40,20.5) 
\item ($J-K$,$J$) = (0.75,11.0) to (0.75,16.5)
\item ($J-K$,$J$) = (0.75,16.5) to (1.70,19.0) 
\end{itemize}

This selection returned a total of 685 low-mass stars and brown dwarfs with
$Z$ magnitude ranging from 12 to 21.5, including known ones recovered by
the GCS (Table \ref{tab_APer:new_members}).
This total number is similar to the number of high probability member 
candidates --- 728 (431) with p\,$>$\,40 (60)\% --- in \APer{} identified 
via the probabilistic approach.

%
%
\subsection{Search for lower mass members}
\label{new_cand_lower_mass}

In this section we search for fainter and cooler substellar members in \APer{}
by dropping the constraint on the $Z$-band detection and later the $Z+Y$ bands.

\subsubsection{$YJHK$ detections}

To extend the \APer{} cluster sequence to fainter brown dwarfs and cooler
temperatures, we searched for potential candidate members undetected in $Z$.
We imposed similar photometric and astrometric criteria as those detailed in 
Section \ref{APer:new_cand_phot_PM} but analysed $Z$ drop--outs as follows:
\begin{itemize}
\item $Y$ $\geq$ 18 and $J$ $\leq$ 19.1 mag
\item Candidates should lie above the line defined by ($Y-J$,$Y$) = (0.55,16.0) 
and (1.40,20.5)
\item Candidates should lie above the line defined by ($J-K$,$J$) = (0.75,16.5) 
and (1.70,19.0)
\item The position on the proper motion vector point diagram of each candidate 
should not deviate from the assumed cluster proper motion by more than 3$\sigma$
\end{itemize}
This selection returned 13 additional member candidates in \APer{}
(Table \ref{tab_APer:YJHK_JHK_detections}). All but four of them are indeed 
undetected in the $Z$-band images and look well detected in the other bands 
after checking the GCS DR9 images. Thus we are left with nine bona-fide
member candidates.

\subsubsection{$JHK$ detections}

We repeated the procedure described above looking for $Z$ and $Y$ non detections.
We additionally applied the following criteria:
\begin{itemize}
\item $J$ = 18--19.1 mag
\item Candidates should lie above the line defined by ($J-K$,$J$) = (0.75,16.5)
and (1.70,19.0)
\item The position on the proper motion vector point diagram of each candidate 
should not deviate from the assumed cluster proper motion by more than 3$\sigma$
\end{itemize}

This query returned 36 new candidate members in \APer{}. After checking the 
GCS images, we retained only eight of them as bona-fide candidates because the 
others actually appear in the $Z$ and/or $Y$ images (although detections are 
not reported in the GCS DR9 catalogue) or have no $Z$ and/or $Y$ images. The 
reasons for the rejection of 28 of the 36 candidates is given in the last column 
of Table \ref{tab_APer:YJHK_JHK_detections}.

%
%
%
\begin{figure}
   \includegraphics[width=\linewidth]{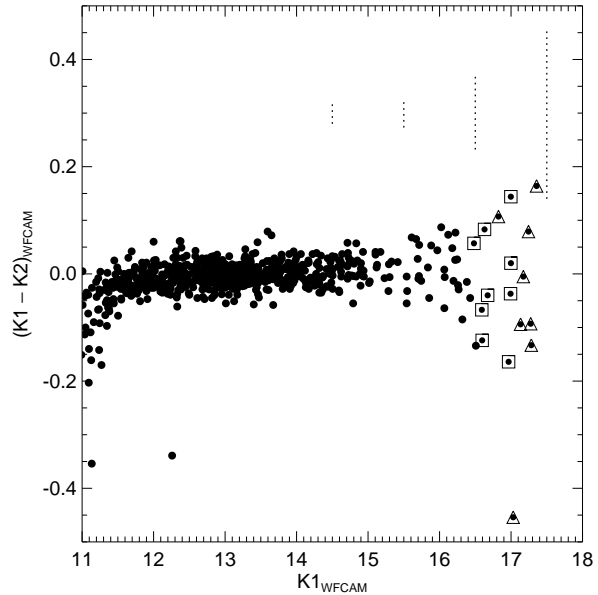}
   \caption{Difference in the $K$ magnitude ($K$1--$K$2) as a function of the 
$K$1 magnitude for all member candidates in \APer{} selected with method \#2\@.
The $YJHK$ and $JHK$-only detections have been added too (dots with open squares 
and open triangles, respectively).
Typical error bars on the $K$1--$K$2 colours as a function of magnitude
are displayed as dotted lines.
} 
   \label{fig_APer:diagram_variability}
\end{figure}
%

%
%
\section{Estimation of the contamination}
\label{APer:contamination}

In this section we estimate the level contamination present in our photometric
and astrometric selection (method \#2).

\begin{figure}
  \includegraphics[width=\linewidth]{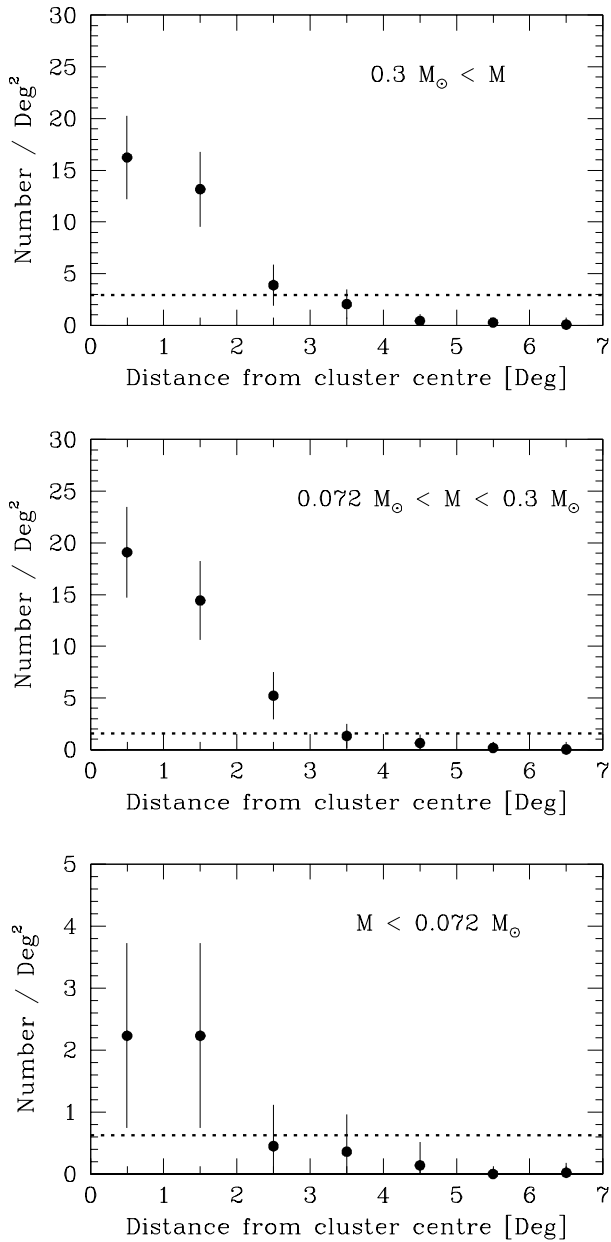}
  \caption{Radial density plots of our
    candidate members of $\alpha$~Per in three mass ranges: above
    0.3 M$_\odot$ (top panel), between 0.72 and 0.3 M$_\odot$
    (middle panel), and below the stellar/substellar limit at
    0.072 M$_\odot$ (low panel). The error bars on each datapoint are
    Poissonian arising from the number of objects in each bin. The
    dotted horizontal line is the estimated contamination per square
    degree for each mass range.}
    \label{fig_APer:radial-plot}
\end{figure}

The number density of field objects in our final list of candidates as
a function of mass is obtained in a similar way as in Boudreault et al.\ (2012, submitted to MNRAS).
We obtained the radial profile of our cluster candidates in three mass ranges: 
above 0.3\,M$_\odot$, between 0.072 and 0.3 M$_\odot$, and below the 
hydrogen-burning limit at 0.072 M$_\odot$ (Fig.\ \ref{fig_APer:radial-plot}).

However, considering the incomplete coverage of the UKIDSS GCS DR9
towards \APer{} (holes present in the coverage due to quality control, 
see Fig.\ \ref{fig_APer:GCScoverage}), all datapoints must be considered as
lower limits: we are only partly complete up to the tidal radius of
\APer{} at 2.91$^{\circ}$ \citep[9.7 pc;][]{makarov06} and up
to 3.5$^{\circ}$ (95\% complete in coverage). Consequently, the estimated
contamination represent an upper limit.

%
%

We used only the number of objects between 3 and
3.5$^{\circ}$ (outside the estimated tidal radius) at each mass range
to obtain an upper limit of contamination. This gives 2.92 objects per
square degree for candidates with masses above 0.3 M$_\odot$, 1.57
between 0.072 and 0.3 M$_\odot$, and 0.62 objects per square degree
for our substellar candidates. Within 3$^{\circ}$ from the cluster
centre, this gives a contamination of 35.1\%, 15.9\% and 50.6\%
for the same mass range respectively, or 26.3\% for the whole
\APer{} sample within 3$^{\circ}$. {\bf{This level of contamination 
brings into agreement within a factor of two the luminosity functions
derived from both selection methods highlighted in this paper
(left-hand sidepanel of Fig.\ \ref{fig_APer:LF_and_MF}).}}

These numbers appear quite large. We stress again that these are upper
limits, since the coverage is not complete. However, we can claim an
completeness higher than 90\% for our cluster candidate list and the 
determination of our mass function. This is justified by the fact that our 
astrometric selection includes all objects within 3$\sigma$ of the cluster's 
mean proper motion (completeness of $>$99\%) and that the lines used in 
our photometric selection go at least 2$\sigma$ bluer from the cluster main 
sequence in all the colour-magnitude diagrams used for the photometric 
selection (completeness of $\sim$95.4\%).

Most of the contaminants of our cluster candidates with masses above
0.1 M$_\odot$ would be Galactic disk late-type and giant stars, while
most of the contaminants of candidates less massive than
0.1 M$_\odot$ would include Galactic disk late-type and giant stars,
but also unresolved galaxies.

%
%
\section{Variability at 90 Myr}
\label{APer:variability}

In this section we discuss the $K$--band variability of the low-mass stars and brown 
dwarfs in \APer{} using the two epochs provided by the GCS\@. First we
considered the candidates extracted with method \#2\@, several of them
being already published in the literature (Tables \ref{tab_APer:early_DR9}).

Figure \ref{fig_APer:diagram_variability} shows ($K$1--$K$2) versus $K$1 
for all candidate members in \APer{} from method \#2\@. The
brightening in the $K$1 = 11--12 mag range is due to the difference in depth
between the first and second epoch, around 0.5 mag both in the
saturation and completeness limit. This is understandable because the exposure
times have been doubled for the second epoch with relaxed constraints on the
seeing requirement and weather conditions. We excluded those objects from
our variability study. Overall, the sequence is very 
well-defined and very few objects appear variable in the $K$-band.

We selected variable objects by looking at the standard deviation, robustly estimated 
as 1.48$\times$ the median absolute deviation which is the median of the 
sorted set of absolute values of deviation from the central value of
the $K$1--$K$2 colour. We identified one potential variable object in the
$K$1 = 11--12 and 12--13 mag range with differences in the $K$-band larger 
than 3$\sigma$ above the standard deviation. No additional variable
source was picked up beyond 3$\sigma$ down to $K$1 = 16.5 mag. The 
candidate selected in the brightest bin appears saturated in the second 
epoch image, suggesting that the variability may be caused by the inaccurate 
photometry derived from saturated sources. The other source does not look 
saturated: its variability may be attributed to the presence of a faint 
companion located south-east at $\sim$1.2 arcsec best visible in the 
$K$2 image due to the greater depth of the second epoch. 
This variability analysis is not feasible for $K$1 $\geq$ 16.5 mag due to 
the small number of \APer{} candidate members beyond that magnitude range.

We conclude that the level of variability at 90 Myr is small, with standard
deviations in the 0.06--0.09 mag range, suggesting that it cannot account for 
the dispersion in the cluster sequence. The same conclusions are drawn from
the high probability sample and are consistent with our analysis of the
Pleiades \citep{lodieu12a} and Praesepe (Boudreault et al.\ subm) samples
although we should point out that a handful of members are found to be variable.

%
%
\begin{table*}
  \caption{Values for the luminosity and mass functions (both in linear and
logarithmic scales) per magnitude and mass bin for the \APer{} open cluster 
from the probabilistic approach. We assumed a distance of 172.4 pc and employed
the NextGen and DUSTY 90 Myr theoretical isochrones. 
}
  \label{tab_APer:LF_MF_probabilistic}
  \begin{tabular}{c c c | c c c | c c c | c c c}
  \hline
Mag range & Mass range  & Mid-mass & dN   & errH & errL & dN/dM & errH & errL & dN/dlogM & errH & errL \cr
  \hline
12.0--12.5 & 0.7380--0.6420 & 0.6900 &   6.01 &   3.60 &   2.40 &   62.60 &   37.50 &   25.00 &  2.00 & 0.47 & 0.51 \\
12.5--13.0 & 0.6420--0.5750 & 0.6085 &  19.49 &   5.50 &   4.39 &  290.90 &   82.07 &   65.47 &  2.61 & 0.25 & 0.25 \\
13.0--13.5 & 0.5750--0.5070 & 0.5410 &  48.33 &   8.01 &   6.93 &  710.74 &  117.73 &  101.97 &  2.95 & 0.15 & 0.15 \\
13.5--14.0 & 0.5070--0.4200 & 0.4635 &  63.98 &   9.05 &   7.98 &  735.40 &  103.97 &   91.76 &  2.89 & 0.13 & 0.13 \\
14.0--14.5 & 0.4200--0.3260 & 0.3730 &  66.16 &   9.18 &   8.12 &  703.83 &   97.66 &   86.37 &  2.78 & 0.13 & 0.13 \\
14.5--15.0 & 0.3260--0.2440 & 0.2850 &  89.67 &  10.51 &   9.46 & 1093.54 &  128.16 &  115.32 &  2.85 & 0.11 & 0.11 \\
15.0--15.5 & 0.2440--0.1830 & 0.2135 &  77.31 &   9.84 &   8.78 & 1267.38 &  161.23 &  143.91 &  2.79 & 0.12 & 0.12 \\
15.5--16.0 & 0.1830--0.1390 & 0.1610 &  70.11 &   9.42 &   8.36 & 1593.41 &  214.04 &  189.96 &  2.77 & 0.13 & 0.13 \\
16.0--16.5 & 0.1390--0.1085 & 0.1237 &  51.75 &   8.25 &   7.18 & 1696.72 &  270.35 &  235.29 &  2.68 & 0.15 & 0.15 \\
16.5--17.0 & 0.1085--0.0869 & 0.0977 &  50.93 &   8.19 &   7.12 & 2357.87 &  379.11 &  329.58 &  2.72 & 0.15 & 0.15 \\
17.0--17.5 & 0.0869--0.0703 & 0.0786 &  19.14 &   5.46 &   4.35 & 1153.01 &  328.90 &  261.82 &  2.32 & 0.25 & 0.26 \\
17.5--18.0 & 0.0703--0.0591 & 0.0647 &   8.26 &   4.00 &   2.83 &  737.50 &  357.29 &  252.70 &  2.04 & 0.40 & 0.42 \\
18.0--18.5 & 0.0591--0.0514 & 0.0553 &   8.79 &   4.09 &   2.92 & 1141.56 &  531.00 &  379.52 &  2.16 & 0.38 & 0.40 \\
18.5--19.0 & 0.0514--0.0459 & 0.0486 &   6.50 &   3.69 &   2.50 & 1181.82 &  671.38 &  454.55 &  2.12 & 0.45 & 0.49 \\
  \hline
\end{tabular}
\end{table*}
%
%
%
\begin{table*}
  \caption{Same as Table \ref{tab_APer:LF_MF_probabilistic} but for the
sample identified with method \#2\@.
}
  \label{tab_APer:LF_MF_method2}
  \begin{tabular}{c c c | c c c | c c c | c c c}
  \hline
Mag range &  Mass range  & Mid-mass & dN   & errH & errL & dN/dM & errH & errL & dN/dlogM & errH & errL \cr
  \hline
12.0--12.5 & 0.7380--0.6420 & 0.6900 &   4.00 &   3.18 &   1.94 &   41.67 &   33.12 &   20.17 &  1.82 & 0.58 & 0.66 \\
12.5--13.0 & 0.6420--0.5750 & 0.6085 &  28.00 &   6.36 &   5.27 &  417.91 &   94.95 &   78.62 &  2.77 & 0.20 & 0.21 \\
13.0--13.5 & 0.5750--0.5070 & 0.5410 &  61.00 &   8.86 &   7.79 &  897.06 &  130.27 &  114.62 &  3.05 & 0.14 & 0.14 \\
13.5--14.0 & 0.5070--0.4200 & 0.4635 &  78.00 &   9.87 &   8.82 &  896.55 &  113.50 &  101.35 &  2.98 & 0.12 & 0.12 \\
14.0--14.5 & 0.4200--0.3260 & 0.3730 &  79.00 &   9.93 &   8.87 &  840.43 &  105.64 &   94.41 &  2.86 & 0.12 & 0.12 \\
14.5--15.0 & 0.3260--0.2440 & 0.2850 & 104.00 &  11.23 &  10.19 & 1268.29 &  137.01 &  124.22 &  2.92 & 0.10 & 0.10 \\
15.0--15.5 & 0.2440--0.1830 & 0.2135 &  97.00 &  10.89 &   9.84 & 1590.16 &  178.47 &  161.25 &  2.89 & 0.11 & 0.11 \\
15.5--16.0 & 0.1830--0.1390 & 0.1610 &  68.00 &   9.29 &   8.23 & 1545.45 &  211.17 &  187.07 &  2.76 & 0.13 & 0.13 \\
16.0--16.5 & 0.1390--0.1085 & 0.1237 &  52.00 &   8.26 &   7.19 & 1704.92 &  270.92 &  235.86 &  2.68 & 0.15 & 0.15 \\
16.5--17.0 & 0.1085--0.0869 & 0.0977 &  42.00 &   7.54 &   6.46 & 1944.44 &  349.00 &  299.14 &  2.64 & 0.17 & 0.17 \\
17.0--17.5 & 0.0869--0.0703 & 0.0786 &  26.00 &   6.17 &   5.07 & 1566.27 &  371.81 &  305.69 &  2.45 & 0.21 & 0.22 \\
17.5--18.0 & 0.0703--0.0591 & 0.0647 &   9.00 &   4.12 &   2.96 &  803.57 &  368.08 &  264.11 &  2.08 & 0.38 & 0.40 \\
18.0--18.5 & 0.0591--0.0514 & 0.0553 &   7.00 &   3.78 &   2.60 &  909.09 &  491.41 &  337.41 &  2.06 & 0.43 & 0.46 \\
18.5--19.0 & 0.0514--0.0459 & 0.0486 &   8.00 &   3.96 &   2.78 & 1454.55 &  719.64 &  506.16 &  2.21 & 0.40 & 0.43 \\
19.0--19.5 & 0.0459--0.0408 & 0.0433 &   7.00 &   3.78 &   2.60 & 1372.55 &  741.94 &  509.43 &  2.14 & 0.43 & 0.46 \\
19.5--20.0 & 0.0408--0.0369 & 0.0389 &  11.00 &   4.43 &   3.28 & 2820.51 & 1135.34 &  840.70 &  2.40 & 0.34 & 0.35 \\
20.0--20.5 & 0.0369--0.0331 & 0.0350 &   3.00 &   2.94 &   1.66 &  789.47 &  772.76 &  436.40 &  1.80 & 0.68 & 0.80 \\
20.5--21.0 & 0.0331--0.0296 & 0.0314 &   1.00 &   2.32 &   0.87 &  285.71 &  663.68 &  247.44 &  1.31 & 1.20 & 2.01 \\
  \hline
\end{tabular}
\end{table*}
%

%
%
\section{The luminosity and initial mass functions}
\label{APer:IMF}

In this section we discuss the cluster luminosity and mass functions derived 
from the samples of member candidates in \APer{} extracted from both methods 
described in the previous section. We did not attempt to correct the mass 
function for binaries, hence, we compare our results to ``system'' mass functions.
Note that contrary to our work in the Pleiades and Praesepe, we are unable 
to estimate the substellar multiplicity due to larger scatter in the single-star
and binary sequences due to crowding.

\subsection{Age and distance of \APer{}}
\label{APer:IMF_age}

Age determinations in open clusters can vary by up to a factor of two
\citep{jeffries01a}: fitting of the upper main-sequence and giant
branch \citep{mermilliod81} comparing with models including some 
convective overshoot \citep{maeder81} tend to yield younger ages than the
lithium test \citep{rebolo92}. In the case of \APer{}, the former method 
gives 51 Myr whereas the latter suggests an age between 85$\pm$10 Myr
\citep{barrado04b} and 90$\pm$10 Myr \citep{stauffer99}. A similar 
discrepancy has been observed for the Pleiades 
\citep[77 Myr vs.\ 120 Myr;][]{mermilliod81,stauffer98}. 
Moreover, \citet{meynet93}
revised the ages of 30 galactic open clusters based on an updated set
of solar-metallicity isochrones (at that time) taking into account mass loss
and moderate overshooting, yielding 52 Myr and 100 Myr for the \APer{} and
the Pleiades, respectively. The latter age for the Pleiades is favoured
by the fitting technique of the main-sequence evolution developed by
\citet{naylor09} which quoted a 68\% confidence interval of 104--117 Myr
(mean value of 115 Myr), in agreement with the careful comparison of model
isochrones to the Pleiades photometric sequence by \citet{bell12}.
Other clusters with ages derived by the lithium depletion boundary tend to
agree with older age estimates although with a possible trend towards slightly
older ages, e.g.\ IC\,4665 \citep[36 Myr vs 28$\pm$4 Myr][]{mermilliod81,manzi08},
IC\,2602 \citep[30--67 Myr vs 46$\pm$6 Myr;][]{kharchenko05a,dobbie10},
NGC\,2547 \citep[20--35 Myr vs 34--36 Myr;][]{naylor02,jeffries05}, or
M35 \citep[200$^{+200}_{-100}$ Myr vs 175 Myr;][]{sung99,barrado01d}.
We will employ the isochrones for the lithium test age of 90 Myr in the case 
of \APer{}, bearing in mind the current uncertainty on its age of the order 
of 10 Myr.

Several distance estimates have been published for \APer{}: 
190.5$^{+7.2}_{-6.7}$ pc by \citet{robichon99}, 
176.2\,$\pm$\,5.0 pc by \citet{pinsonneault98} and \citet{makarov06}.
The latest value derived from a revised reduction of the Hipparcos
data by \citet{vanLeeuwen09} suggests a distance of 172.4$\pm$2.7 pc
which we adopt in this work.

To summarise, we adopt in this work a distance of 172.4 pc 
\citep{vanLeeuwen09} and employed the Lyon group NextGen \citep{baraffe98}
and DUSTY \citep{chabrier00c} models at an age of 90 Myr to convert the 
luminosity function into a mass function. We should point out that
the lowest mass brown dwarfs in \APer{} are warmer than 1400\,K, the
upper limit where the COND models should be used \citep{baraffe02}.

%
%
%
\begin{figure*}
   \includegraphics[width=0.49\linewidth]{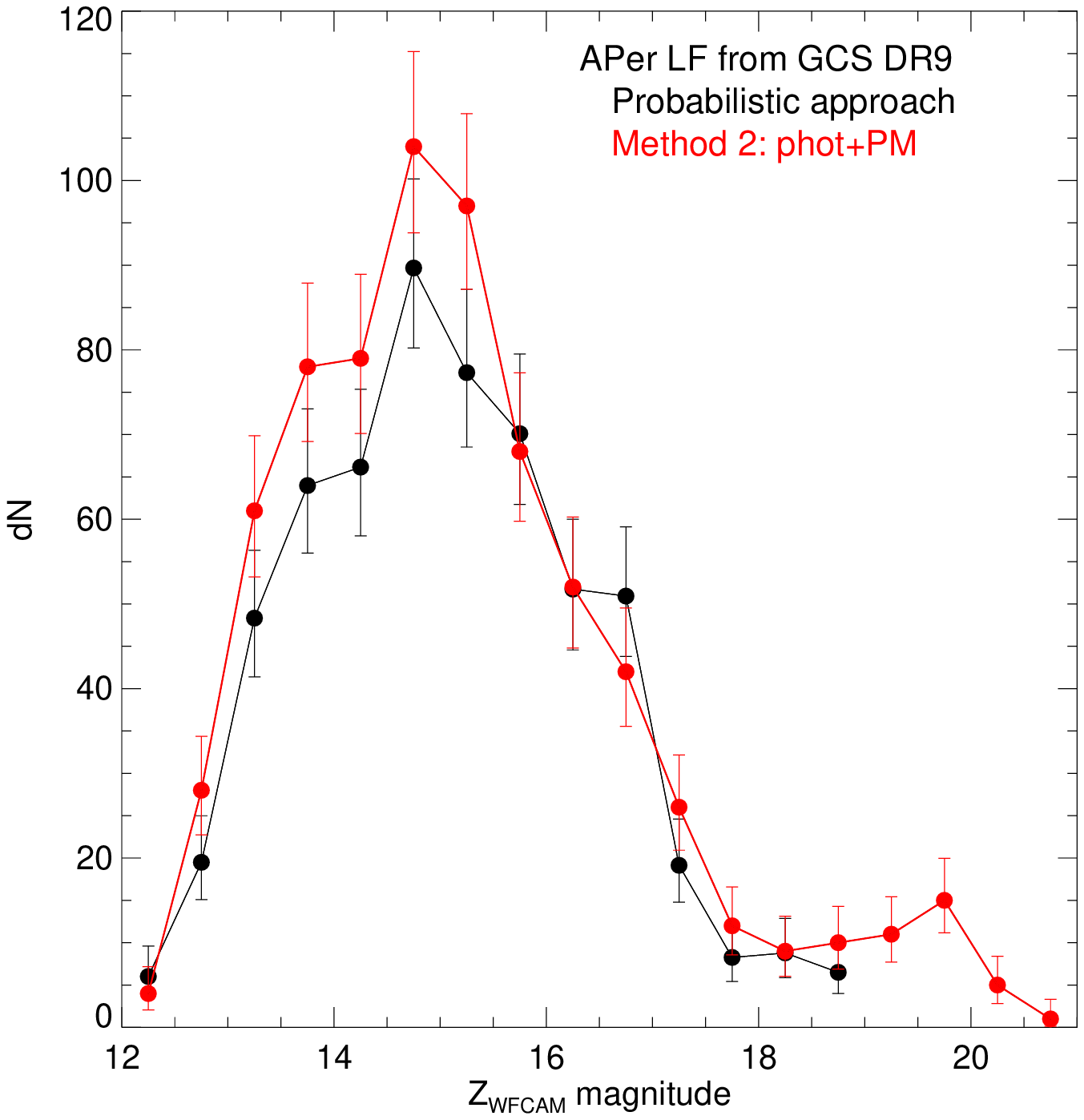}
   \includegraphics[width=0.49\linewidth]{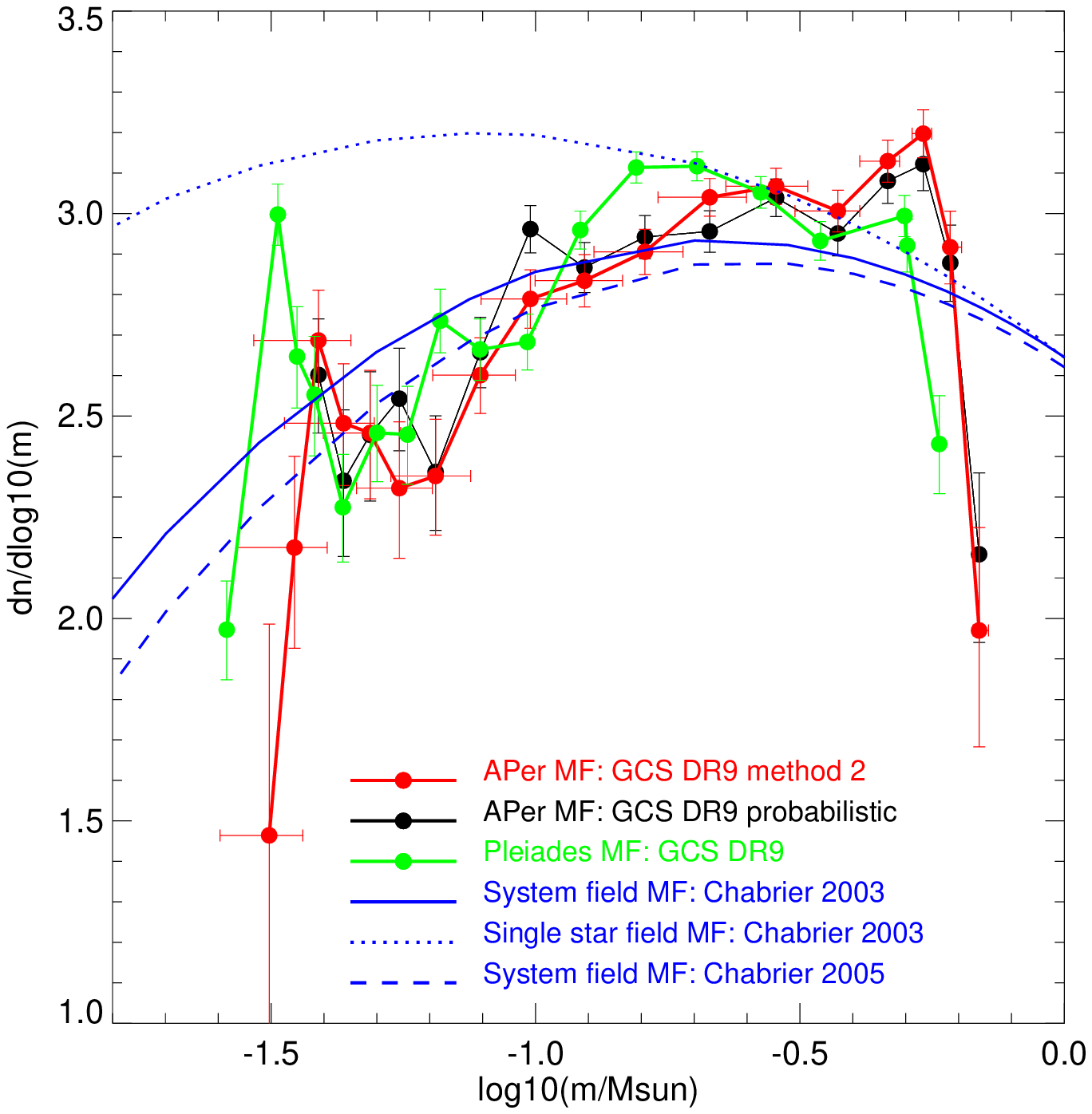}
   \caption{Luminosity (left) and system mass (right) functions derived from 
our analysis of the UKIDSS GCS DR9 sample of member candidates in \APer{}. 
Error bars are Gehrels errors. The brightest bin and the last bins are
very likely contaminated because of saturation and incompleteness, 
respectively. The left-hand side panel compares the luminosity
function obtained from the probabilistic approach (black symbols and black 
line) and the luminosity function derived from the selection outlined by 
method \#2 (red colour). Note that the sample of method \#2 extends two magnitude
bins fainter but they are incomplete as is the brightest bin due to saturation. 
The right-hand side panel compares the \APer{} mass 
function derived from this probabilitic approach (filled black dots linked 
by a solid line) and the mass function derived from method \#2 (red symbols 
and red line). Error bars on the mass (x-axis) are 3$\sigma$ uncertainties
considering the errors on the age and distance of \APer{}. The Pleiades 
mass function derived in a similar manner is overplotted in green for 
comparison along with the field (system) mass functions in blue 
\citep{chabrier05a}. 
} 
   \label{fig_APer:LF_and_MF}
\end{figure*}
\subsection{The luminosity function}
\label{APer:LF}

In this section, we construct two luminosity functions: i) we used the sample
of 10,176 stars in \APer{} with computed membership probabilities
(Section \ref{APer:new_cand_probabilistic}); and ii), the 685
candidates identified with method \#2 (Section \ref{APer:new_cand_phot_PM}).
The luminosity function of the former method is derived by summing membership
probabilities of all stars fitted to distribution functions in the vector
point diagram, whereas the luminosity function of the latter is derived
simply by summing the number of member candidates.

Both luminosity functions i.e.\ the number of stars and brown dwarfs as a
function of magnitude plotted per 0.5 mag bin is displayed in
Fig.~\ref{fig_APer:LF_and_MF}. Both luminosity functions look very similar 
and match each other within the error bars. The numbers of objects per 0.5 mag
bin increase quickly to reach a peak around $Z$ = 14.5--15 and drop off 
afterwards down to the completeness of our survey with a possible peak around
$Z$ = 19.5--20 mag (Tables \ref{tab_APer:LF_MF_probabilistic} \& 
\ref{tab_APer:LF_MF_method2}). The brightest bin is a lower limit 
due to the saturation limit of the GCS survey. The last four bins included in
method \#2 are not present in the probabilistic approach because the broad 
cluster distribution and low separation from the field causes the 
probabilities to be washed out. All bins in the probabilistic luminosity
function are complete while the 
last two bins from method \#2 are incomplete due to the constraints imposed
on the $Z$-band detection. Moreover, the \APer{} luminosity function is very 
similar to the Pleiades one derived in a similar manner using the same 
homogeneous survey \citep{lodieu12a} although less populated mainly because
of the smaller areal coverage.

\subsection{The mass function}
\label{APer:MF}

In this section we adopt the logarithmic form of the Initial Mass Function as
originally proposed by \citet{salpeter55}:
$\xi$($\log_{10}m$) = d$n$/d$\log_{10}$($m$) $\propto$ m$^{-x}$ where
the exponent of the mass spectrum $\alpha = x + 1$ following the formulation 
of \cite{chabrier03}.
The $Z$ = 12--21 mag range translates into masses between $\sim$0.74 and 
$\sim$ 0.03 M$_{\odot}$ (19 mag and 0.046 M$_{\odot}$ in the case of the
probabilistic approach), assuming a revised distance of 172.4 pc 
\citep{vanLeeuwen09} and an age of 90 Myr for which the models are computed.

We included in Fig.\ \ref{fig_APer:LF_and_MF} errors in both the
x-axis ($\log$M) and y-axis (dN/d$\log$M) as follows. For the error bars
on the masses, we considered three times the uncertainties on the age
\citep[90$\pm$10 Myr;][]{stauffer99} and distance
\citep[172.4$\pm$2.7 pc;][]{vanLeeuwen09} of \APer{} given us a validity
range of 3$\sigma$ on the x-axis.
Hence, we computed the masses with the 60 Myr NextGen and DUSTY isochrones
shifted at a distance of 164.3 pc to define the lower limit and repeated the
procedure with the 120 Myr isochrones for a distance of 180.5 pc as upper limits.
The uncertainties on the y-axis i.e.\ the dN/d$\log$M values are simply
Gehrels error bars. This \APer{} mass function, directly compared to the 
Pleiades \citep{lodieu12a} and the field \citep{chabrier05a} mass functions, 
agree within the error bars. We should point out the recent mass function 
of the field published by \citet{kroupa11} and described as a power-law is 
almost identical to the log-normal form of \citet{chabrier05a}.

\begin{figure}
  \includegraphics[width=\linewidth,angle=270,scale=0.8]{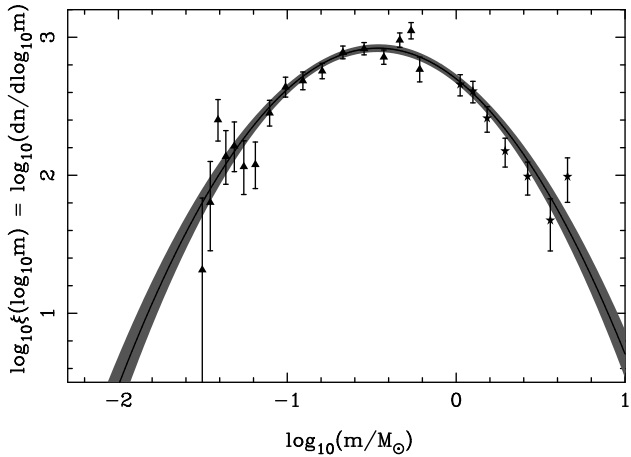}
   \caption{Log-normal fit to the GCS DR9 data (method~\#2; triangles with error bars) in
   conjunction with higher mass data points (stars with error bars) taken from
   Prosser (1992). The least--squares fit to the data points is the solid line
   with the shaded region corresponding to a formal $1\sigma$ uncertainty.} 
 \label{fig_APer:AlphaPerlognormplot}   
\end{figure}

In Fig.~\ref{fig_APer:AlphaPerlognormplot} we show a log-normal fit for \APer{} 
incorporating higher mass data points from \citet{prosser92} in order to
provide constraint on the parameters of the fit, in particular the 
characteristic mass which requires sufficient points on both sides of the peak 
in the function. We translated the `corrected' luminosity function values 
making a small update 
to the absolute magnitudes for the distance modulus used here (6.18) over the 
value of 6.1 in \citet{prosser92}. The visual band mass--luminosity relation 
used comes from \citet{marigo08} evolutionary 
models\footnote{http://stev.oapd.inaf.it/cgi-bin/cmd\_2.3}.
We include in the fit only those higher mass points that are complete, i.e.\ for
M$_{\rm V}<+5$ from \citet{prosser92}, and excluding our own highest mass point 
from the GCS luminosity function, but we include the four lowest mass points 
from the GCS since excluding them does not significantly alter the fit.
The mass function appears to be well represented by a log-normal with 
goodness--of--fit $\chi^2_{\nu}\approx2.3$ which indicates some systematic 
fluctuations over and above the assumed sampling errors that could easily be 
due to sample contamination and/or systematic errors resulting from the 
assumed models.

%
%

\begin{figure}
  \includegraphics[width=\linewidth,scale=0.8]{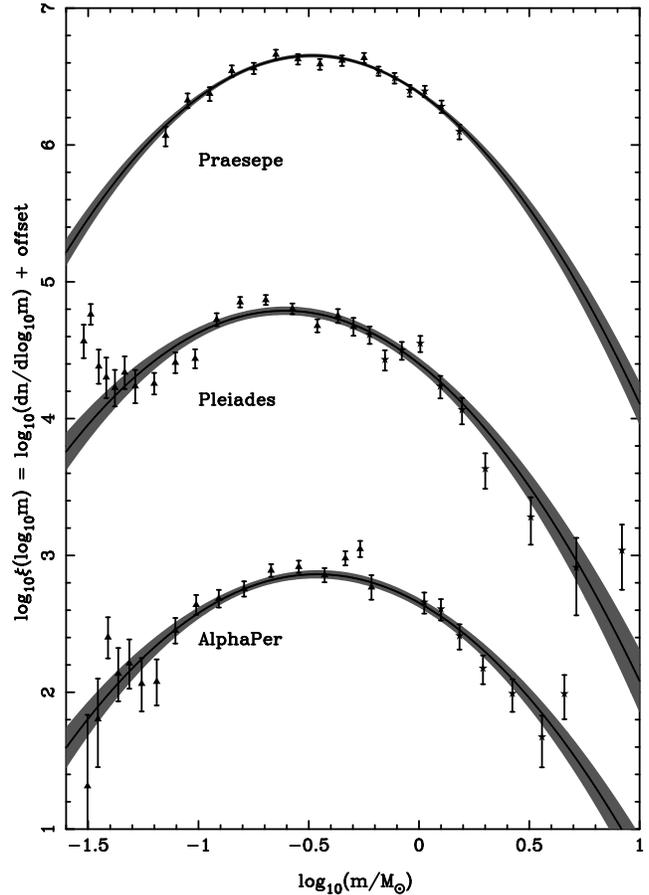}
   \caption{Log-normal fit to the GCS DR9 data (triangles with error bars) in
   conjunction with higher mass data points (stars with error bars) taken for
   the Pleiades (Lodieu et al.\ 2012a excluding the 3 lowest mass bins; higher 
   mass points from the unpublished compilations of Prosser and Stauffer, see 
   for example \citet{hambly99} and references therein); Alpha Per (this work); 
   and Praesepe (Boudreault et al.\ subm to MNRAS; higher
   mass points from \citet{jones91}). In each case,
   least--squares fits to the data points are the solid line
   with the shaded region corresponding to a formal $1\sigma$ uncertainty.   
 \label{fig_APer:lognormplot3}   
   } 
\end{figure}

It is interesting to compare this mass function with those 
from the Pleiades \citep{lodieu12a} and Praesepe
(Boudreault et al.\ subm.\ to MNRAS), with similar higher mass constraints 
from optical photographic plate surveys -- see Fig.~\ref{fig_APer:lognormplot3}.
In the case of the Pleiades and Alpha Per, the higher mass luminosity functions 
have been taken as complete and no normalisation has been performed relative 
to the GCS luminosity functions, whereas for Praesepe we found that the mass
function resulting from the \citet{jones91} luminosity function is 
discontinuous with the GCS mass function from Boudreault at al.\ (2012, subm
to MNRAS). We determined a relative normalisation of 0.447 in the log 
(a factor 2.8) for a minimal chi-squared in the log-normal fit for Praesepe.

%
%

In Table~\ref{tab:lognormpars} we compare the log-normal fit parameters to the 
field system mass function parameterised by \citet{chabrier03} and 
\citet{chabrier05a}. There is some marginal evidence here for a variation in 
characteristic mass at the 1 to 2$\sigma$ level between \APer{} and Praesepe 
and the Pleiades, but this must be treated with caution given 
the range of goodness--of--fits obtained ($1.0<\chi^2_\nu<4.4$) and 
particularly the significant departure from the fit for the Pleiades at
the low mass end. There is a clear statistically significant difference between 
the dispersion values of the field and \APer{} mass functions, not unexpected
due to the difference in age. While we caution that the 
fitted values can be sensitive to the relative normalisation between the GCS 
and higher mass data, changes in the relative offsets tend to narrow the 
log-normal fit rather than broaden it. In any case, it is interesting to note 
the general log-normal trend in these wide mass range mass functions.

%
%
%
\begin{table}
\begin{center}
\begin{tabular}{c c c c}
  \hline
Population & Characteristic & Dispersion & $\chi^2_\nu$ \\
           & mass $m_{\rm C}$ (M$_{\odot}$) & $\sigma$ & \\
  \hline
  \multicolumn{4}{c}{ }\\
Alpha Per & $0.344\pm0.045$ & $0.458\pm0.019$ & 2.275 \\
Pleiades  & $0.247\pm0.047$ & $0.456\pm0.023$ & 4.382 \\
Praesepe  & $0.328\pm0.035$ & $0.434\pm0.015$ & 0.962 \\
  \multicolumn{4}{c}{ }\\
  \hline
Field (Chabrier 2003)     & 0.22 & 0.57 & \\
Field (Chabrier 2005)     & 0.25 & 0.55 & \\
  \hline
\end{tabular}
\caption{Comparison between log-normal mass function parameters for the Alpha 
Per, Pleiades and Praesepe clusters as determined from GCS DR9 data in 
conjunction with higher mass bin data from optical photographic proper motion 
surveys, compared with the field system mass function parameters quoted by 
\citet{chabrier03} and \citet{chabrier05a}.}
  \label{tab:lognormpars}
\end{center}
\end{table}

Assuming that the observed lithium depletion boundary is at
M $\sim$ 0.075 M$_{\odot}$ \citep[M$_{Z}$ = 11.155;][]{stauffer99,barrado04b}
and a distance of 172.4 pc, the sample extracted by method \#2 contains
685 \APer{} member candidates, divided up into 632 stars (92.3\%) and
53 brown dwarfs (7.7\%). Lower percentages of brown dwarfs are obtained 
considering the sample of 431--728 high probability members (p$\geq$40--60\%)
identified in the probabilistic approach, because of larger uncertainties on
the probabilities at the faint end. Hence, the star 
($\sim$0.6--0.08 M$_{\odot}$) to brown dwarf (0.08--0.04 M$_{\odot}$)
ratio in \APer{} spans 11.9 (10.4--12.7; 3$\sigma$ limits using the lower 
and upper distance estimates) to 16.8$^{-2.0}_{+2.5}$--33.3$_{-1.9}^{+4.6}$, 
in agreement with measurements
in IC\,348 \citep[8.3--11.6;][]{luhman03b,andersen06} but higher than
other open clusters like M35 \citep[4.5;][]{barrado01a} or the Pleiades
\citep[3.7 and 5.7--8.8;][]{bouvier98,lodieu12a}, young star-forming regions 
\citep[3.0--6.4 for the Trapezium Cluster; 3.8-4.3 for $\sigma$ Orionis; 3.8 
for Chamaeleon;][]{hillenbrand00,muench02,andersen06,luhman07d,lodieu09e},
the field \citep[1.7--5.3;][]{kroupa02,chabrier05a,andersen06}, and
hydrodynamical simulations of star clusters \citep[3.8--5.0;][]{bate09,bate12}.
We list the ranges of the ratios because the stellar and substellar
intervals differ slightly from study to study.

%
%
\section{Summary}
\label{APer:summary}

We have presented the outcome of a wide ($\sim$56 square degrees) and deep 
($J$ $\sim$ 19.1 mag) survey in the \APer{} open cluster as part of the UKIDSS 
Galactic Clusters Survey Data Release 9\@. The main results of our study
can be summarised as follows:
\begin{itemize}
\item we recovered member candidates in \APer{} previously published
and updated their membership assignations
\item we selected photometrically and astrometrically potential \APer{} 
member candidates using two independent but complementary methods: the
probabilistic analysis and a more standard method combining photometry and 
proper motion cuts
\item we investigated the $K$-band variability of \APer{} cluster members and
found virtually no variability at the level of 0.06--0.09 mag
\item we derived the luminosity function from both selection methods and
found no difference within the error bars
\item we derived the \APer{} mass function over the 0.5--0.04 M$_{\odot}$
mass range: its shape is similar to the Pleiades mass function and best 
represented by a log-normal form with a characteristic mass of 0.34 M$_{\odot}$
and a dispersion of 0.46\@.
\end{itemize}

This paper represents a significant improvement in our census of the 
\APer{} low-mass and substellar population as well as our knowledge of
the mass function across the hydrogen-burning limit over the entire cluster.
We believe that this paper will represent a reference for many more years
to come in \APer{}. We will now extend this study to other regions
surveyed by the GCS to address the question of the universality of the
mass function using an homogeneous set of photometric and astrometric data.
Future work to constrain current models of star formation includes the search
for companions to investigate their multiplicity properties, the determination 
of the radial velocities of \APer{} members, and deeper surveys
to test the theory of the fragmentation limit.

%
%
\section*{Acknowledgments}

NL is funded by the Ram\'on y Cajal fellowship number 08-303-01-02 and the 
national program AYA2010-19136 funded by the Spanish ministry of science and 
innovation (SB is also funded by this grant). We thank the referee, Tim
Naylor, for his careful review and for advice concerning cluster age 
determinations which has led to an improved paper.
 This work is based in part on data obtained as part of the UKIRT 
Infrared Deep Sky Survey (UKIDSS). The UKIDSS project is defined in 
\citet{lawrence07}. UKIDSS uses the UKIRT Wide Field Camera 
\citep[WFCAM;][]{casali07}. The photometric system is described in 
\citet{hewett06}, and the calibration is described in \citet{hodgkin09}.
The pipeline processing and science archive are described in Irwin et al.\
(in prep) and \citet{hambly08}, respectively.
We thank our colleagues at the UK Astronomy Technology Centre, the Joint 
Astronomy Centre in Hawaii, the Cambridge Astronomical Survey and Edinburgh 
Wide Field Astronomy Units for building and operating WFCAM and its 
associated data flow system.
We are grateful to France Allard's homepage that we used to download the 
NextGen and DUSTY isochrones at 60, 90, and 120 Myr for the WFCAM filters.

This research has made use of the Simbad database, operated at the Centre de 
Donn\'ees Astronomiques de Strasbourg (CDS), and of NASA's Astrophysics Data 
System Bibliographic Services (ADS). This publication has also made use of data 
products from the Two Micron All Sky Survey, which is a joint project of the
University of Massachusetts and the Infrared Processing and Analysis 
Center/California Institute of Technology, funded by the National Aeronautics 
and Space Administration and the National Science Foundation.

%
%
\bibliographystyle{mn2e}
\bibliography{../../AA/mnemonic,../../AA/biblio_old,nige}

%
%
\appendix

%
%
\section{Table of known member candidates previously published 
in \APer{} and recovered in UKIDSS GCS DR9\@.}

\begin{table*}
  \caption{Sample of known member candidates previously published in \APer{} 
and recovered in GCS DR9\@. We list the equatorial coordinates (J2000), 
GCS $ZYJHK1K2$ photometry, proper motions (in mas/yr) and their errors, 
reduced chi-squared statistic of the astrometric fit for each source 
($\chi^{2}$ value), membership probabilities when available from our 
probabilistic study, and names from the literature.
A $---$ line in the probability column means that the object lacks measurement
\APer{} member candidates are ordered by increasing right ascension.
This table is available electronically in the online version of the journal.
}
  \label{tab_APer:early_DR9}
  \begin{tabular}{@{\hspace{0mm}}c @{\hspace{2mm}}c c @{\hspace{2mm}}c @{\hspace{2mm}}c @{\hspace{2mm}}c @{\hspace{2mm}}c @{\hspace{2mm}}c c @{\hspace{2mm}}c c c l@{\hspace{0mm}}}
  \hline
R.A.\ & Dec.\  &  $Z$  &  $Y$  &  $J$  &  $H$  & $K$1 & $K$2 & $\mu_{\alpha}cos\delta$\,$\pm$\,err & $\mu_{\delta}$\,$\pm$\,err & $\chi^{2}$ & Prob & Name \cr
 \hline
02 58 17.66 & +48 28 00.4 & 16.152 & 15.700 & 15.071 & 14.531 & 14.187 & 14.175 &    23.07$\pm$2.91 &   $-$14.86$\pm$2.91 &  0.59 & ---  & DH12\_Prob73.7 \cr
03 01 21.38 & +48 35 23.3 & 13.971 & 13.664 & 13.142 & 12.494 & 12.257 & 12.267 &    24.39$\pm$2.86 &   $-$21.71$\pm$2.86 &  0.11 & 0.77 & DH15\_Prob70.8 \cr
 \ldots{}   & \ldots{}   & \ldots{} & \ldots{} & \ldots{} & \ldots{} & \ldots{} & \ldots{}  & \ldots{}  & \ldots{} & \ldots{} & \ldots{} & \ldots{} \cr
03 50 37.08 & +48 12 31.4 & 14.124 & 13.621 & 13.079 & 12.534 & 12.234 & 12.226 &    21.45$\pm$2.03 &   $-$35.54$\pm$2.03 &  6.38 & ---  & AP265\_M9.9\_Y? \cr
03 50 37.08 & +48 12 31.4 & 14.124 & 13.621 & 13.079 & 12.534 & 12.234 & 12.226 &    21.45$\pm$2.03 &   $-$35.54$\pm$2.03 &  6.38 & ---  & DH302\_Prob79.1 \cr
 \hline
\end{tabular}
\end{table*}
%

%
%
\section{Table of new member candidates in \APer{} identified in the probabilistic approach}

\begin{table*}
  \caption{Coordinates (J2000), near-infrared ($ZYJHK$1$K$2) photometry, and 
proper motions (in mas/yr) for all high probability (p$\geq$40\%) members in 
\APer{} identified in the UKIDSS GCS DR9 using the probabilistic approach. 
The last column gives the membership probability. Sources are ordered by 
increasing right ascension. This table is available electronically in the 
online version of the journal.
}
  \label{tab_APer:new_highprob}
  \begin{tabular}{@{\hspace{0mm}}c c c c c c c c c c c@{\hspace{0mm}}}
  \hline
R.A.\ & Dec.\  &  $Z$  &  $Y$  &  $J$  &  $H$  & $K$1 & $K$2 & $\mu_{\alpha}cos\delta$ & $\mu_{\delta}$ & Prob \\
 \hline
02 58 52.52 & +49 40 32.6 & 14.543 &  ---   & 13.655 & 12.993 & 12.761 & 12.748 &    26.74 &   $-$22.66 & 0.71 \cr
02 58 57.10 & +50 44 41.4 & 15.074 & 14.759 & 14.213 & 13.590 & 13.335 & 13.344 &    23.90 &   $-$20.86 & 0.61 \cr
\ldots{}  & \ldots{} & \ldots{} & \ldots{} & \ldots{} & \ldots{} & \ldots{} & \ldots{} & \ldots{}  & \ldots{}  & \ldots{} \cr
03 50 01.17 & +48 20 57.3 & 16.587 & 16.104 & 15.490 & 14.812 & 14.494 & 14.462 &    20.58 &   $-$26.20 & 0.46 \cr
03 50 20.08 & +48 13 54.8 & 15.402 & 15.029 & 14.504 & 13.940 & 13.645 & 13.617 &    25.02 &   $-$19.56 & 0.43 \cr
 \hline
\end{tabular}
\end{table*}

%
%
\section{Table of new member candidates in \APer{} selected with method \#2}

\begin{table*}
  \caption{Coordinates (J2000), near-infrared ($ZYJHK$1$K$2) photometry, and 
proper motions (in mas/yr) for all member candidates in \APer{} identified in 
the UKIDSS GCS DR9 with the standard method (method \#2), including known
members from earlier studies. Sources are ordered by increasing right 
ascension. This table is available electronically in the online version of 
the journal.
}
  \label{tab_APer:new_members}
  \begin{tabular}{@{\hspace{0mm}}c c c c c c c c c c@{\hspace{0mm}}}
  \hline
R.A.\ & Dec.\  &  $Z$  &  $Y$  &  $J$  &  $H$  & $K$1 & $K$2 & $\mu_{\alpha}cos\delta$ & $\mu_{\delta}$ \\
 \hline
02 57 51.18 & +48 08 29.0 & 16.810 & 16.101 & 15.536 & 14.900 & 14.598 & 14.632 &    17.56$\pm$2.96 &   $-$29.20$\pm$2.96 \cr
02 57 52.10 & +48 23 58.8 & 17.175 & 16.459 & 15.810 & 15.192 & 14.851 & 14.828 &    22.12$\pm$3.05 &   $-$17.13$\pm$3.05 \cr
\ldots{}  & \ldots{} & \ldots{} & \ldots{} & \ldots{} & \ldots{} & \ldots{} & \ldots{} & \ldots{}  & \ldots{} \cr
03 50 18.91 & +48 24 59.1 & 18.766 & 17.618 & 16.684 & 16.108 & 15.610 & 15.542 &    18.12$\pm$2.54 &   $-$25.55$\pm$2.54 \cr
03 50 35.47 & +47 25 56.3 & 17.188 & 16.424 & 15.716 & 15.156 & 14.728 & 14.730 &    22.78$\pm$2.59 &   $-$29.06$\pm$2.59 \cr
 \hline
\end{tabular}
\end{table*}

%
%
\section{Table of member candidates in \APer{} with $YJHK$ and $JHK$-only detections}

\begin{table*}
  \caption{Coordinates (J2000), near-infrared ($ZYJHK$1$K$2) photometry, and proper 
motions (in mas/yr) for $YJHK$-only (top) and $JHK$-only (bottom) detections.
}
  \label{tab_APer:YJHK_JHK_detections}
  \begin{tabular}{@{\hspace{0mm}}c c c c c c c c c c c@{\hspace{0mm}}}
  \hline
R.A.\ & Dec.\  &  $Z$  &  $Y$  &  $J$  &  $H$  & $K$1 & $K$2 & $\mu_{\alpha}cos\delta$ & $\mu_{\delta}$ & Comments \\
 \hline
03 16 26.24 & +49 00 12.2 &  ---  & 20.314 & 18.797 & 17.849 & 16.993 & 17.030 &    14.76$\pm$4.60 &   $-$40.10$\pm$4.60 &       \cr
03 21 14.97 & +49 14 23.2 &  ---  & 19.548 & 18.220 & 17.410 & 16.673 & 16.713 &    21.93$\pm$3.59 &   $-$18.70$\pm$3.59 &       \cr
03 23 09.75 & +50 20 03.3 &  ---  & 19.319 & 18.017 & 17.307 & 16.591 & 16.658 &    18.62$\pm$4.84 &   $-$36.35$\pm$4.84 &       \cr
03 24 01.62 & +46 48 52.7 &  ---  & 19.291 & 18.038 & 17.273 & 16.596 & 16.720 &    11.99$\pm$5.96 &   $-$14.60$\pm$5.96 &       \cr
03 27 49.28 & +50 42 26.3 &  ---  & 18.483 & 17.404 & 16.771 & 16.191 & 16.127 &    13.76$\pm$4.35 &   $-$21.26$\pm$4.35 & detected in Z \cr
03 28 11.64 & +51 46 50.6 &  ---  & 18.138 & 15.299 & 14.980 & 14.781 & 14.804 &    28.61$\pm$3.04 &   $-$22.53$\pm$3.04 & detected in Z \cr
03 28 38.15 & +48 59 51.1 &  ---  & 20.508 & 18.738 & 17.845 & 16.997 & 16.853 &    26.30$\pm$4.44 &   $-$29.45$\pm$4.44 &       \cr
03 29 49.62 & +48 35 05.3 &  ---  & 20.112 & 18.739 & 17.846 & 16.998 & 16.978 &    17.76$\pm$5.21 &   $-$35.06$\pm$5.21 &       \cr
03 30 52.69 & +50 28 34.7 &  ---  & 19.908 & 18.498 & 17.390 & 16.481 & 16.424 &    28.15$\pm$4.34 &   $-$30.99$\pm$4.34 &       \cr
03 32 27.13 & +48 00 54.3 &  ---  & 19.428 & 18.138 & 17.338 & 16.629 & 16.546 &    31.10$\pm$4.60 &   $-$23.63$\pm$4.60 &       \cr
03 32 42.65 & +50 01 39.8 &  ---  & 20.449 & 19.026 & 17.926 & 16.966 & 17.130 &    20.66$\pm$6.40 &   $-$34.40$\pm$6.40 &       \cr
03 36 03.86 & +50 39 57.7 &  ---  & 20.269 & 18.899 & 17.680 & 17.017 & 17.059 &     5.34$\pm$6.87 &    $-$6.19$\pm$6.87 & detected in Z \cr
03 39 53.40 & +49 06 59.5 &  ---  & 20.228 & 18.785 & 17.865 & 17.098 & 16.986 &     8.58$\pm$7.88 &   $-$22.64$\pm$7.88 & detected in Z \cr
 \hline
02 59 48.86 & +47 50 31.8 &  ---  &  ---  & 18.810 & 17.938 & 17.138 & 17.249 &     4.88$\pm$6.53 &   $-$14.06$\pm$6.53 & no Z,Y images \cr
03 01 14.17 & +49 03 05.5 &  ---  &  ---  & 18.798 & 17.916 & 16.824 & 17.112 &     7.84$\pm$6.33 &   $-$10.54$\pm$6.33 & no Y image,detected in Z? \cr
03 09 07.55 & +49 37 36.8 &  ---  &  ---  & 19.067 & 18.241 & 17.339 & 17.336 &    $-$0.08$\pm$9.06 &    $-$1.19$\pm$9.06 & no Y image \cr
03 10 32.62 & +49 25 19.4 &  ---  &  ---  & 19.043 & 18.290 & 17.289 & 17.416 &     0.63$\pm$8.53 &    $-$5.17$\pm$8.53 & detected in Y \cr
03 11 26.76 & +49 13 52.2 &  ---  &  ---  & 19.016 & 18.315 & 17.295 & 17.414 &     8.31$\pm$8.81 &    $-$1.26$\pm$8.81 & detected in Z+Y \cr
03 12 05.31 & +49 02 16.0 &  ---  &  ---  & 18.975 & 17.868 & 17.244 & 17.283 &     8.17$\pm$8.49 &   $-$12.99$\pm$8.49 & no Y image \cr
03 12 25.76 & +49 43 42.5 &  ---  &  ---  & 19.058 & 18.137 & 16.984 & 17.066 &    14.33$\pm$7.12 &   $-$16.47$\pm$7.12 & no Y image \cr
03 14 56.42 & +50 08 28.3 &  ---  &  ---  & 19.048 & 17.988 & 17.192 & 17.061 &     1.16$\pm$8.21 &   $-$21.09$\pm$8.21 & detected in Z+Y?? \cr
03 16 22.23 & +52 32 00.9 &  ---  &  ---  & 18.591 & 17.299 & 16.577 & 16.743 &    23.79$\pm$7.20 &   $-$10.38$\pm$7.20 & spike of a bright star \cr
03 16 25.02 & +52 32 09.1 &  ---  &  ---  & 18.585 & 17.861 & 17.018 & 17.308 &    32.54$\pm$5.36 &   $-$26.23$\pm$5.36 & detected in \cr
03 17 37.31 & +47 05 14.6 &  ---  &  ---  & 18.988 & 18.136 & 17.167 & 17.125 &    16.59$\pm$8.48 &    $-$7.33$\pm$8.48 & no Y image \cr
03 17 49.13 & +46 58 35.3 &  ---  &  ---  & 19.058 & 18.136 & 17.290 & 17.220 &     5.62$\pm$9.28 &    $-$3.97$\pm$9.28 & detected in Y \cr
03 18 23.96 & +46 26 49.6 &  ---  &  ---  & 18.573 & 17.553 & 16.699 & 16.774 &     6.39$\pm$5.51 &   $-$42.34$\pm$5.51 & detected in Y \cr
03 19 11.02 & +51 24 47.0 &  ---  &  ---  & 19.093 & 18.463 & 17.314 & 17.329 &     2.22$\pm$9.51 &    $-$9.80$\pm$9.51 & no Y image \cr
03 19 19.19 & +46 10 18.8 &  ---  &  ---  & 19.001 & 17.768 & 16.921 & 16.975 &    28.50$\pm$9.56 &   $-$13.52$\pm$9.56 & detected in Z+Y \cr
03 20 41.79 & +50 45 38.6 &  ---  &  ---  & 18.796 & 18.187 & 17.030 & 17.484 &    $-$0.96$\pm$9.46 &   $-$10.49$\pm$9.46 &       \cr
03 21 14.74 & +46 36 27.1 &  ---  &  ---  & 19.007 & 18.143 & 17.180 & 17.220 &     3.95$\pm$9.83 &   $-$18.58$\pm$9.83 & detected in Z+Y?? \cr
03 21 53.44 & +46 47 02.6 &  ---  &  ---  & 19.092 & 18.201 & 17.357 & 17.193 &     6.41$\pm$10.48 &   $-$12.39$\pm$10.48 &       \cr
03 23 02.14 & +52 13 58.8 &  ---  &  ---  & 18.945 & 17.873 & 17.000 & 17.120 &    11.70$\pm$10.54 &    $-$6.76$\pm$10.54 & detected in Y \cr
03 24 03.07 & +50 03 01.0 &  ---  &  ---  & 19.098 & 17.911 & 17.071 & 17.090 &    16.35$\pm$7.93 &   $-$20.52$\pm$7.93 & detected in Y \cr
03 24 32.00 & +47 04 29.5 &  ---  &  ---  & 18.656 & 17.941 & 17.035 & 16.878 &     5.29$\pm$8.02 &   $-$11.42$\pm$8.02 & detected in Z+Y \cr
03 24 46.24 & +46 36 25.4 &  ---  &  ---  & 19.040 & 18.289 & 17.244 & 17.165 &    10.75$\pm$10.39 &   $-$21.22$\pm$10.39 &       \cr
03 26 14.34 & +51 55 36.4 &  ---  &  ---  & 18.718 & 17.778 & 16.824 & 16.717 &     2.88$\pm$8.33 &    $-$7.91$\pm$8.33 &       \cr
03 27 14.93 & +52 15 58.3 &  ---  &  ---  & 18.750 & 17.710 & 16.717 & 16.713 &     6.39$\pm$6.88 &    $-$9.24$\pm$6.88 & detected in Z+Y \cr
03 27 32.27 & +47 11 45.4 &  ---  &  ---  & 19.079 & 18.325 & 17.133 & 17.227 &    $-$1.16$\pm$10.13 &   $-$16.30$\pm$10.13 &       \cr
03 27 43.73 & +46 55 02.9 &  ---  &  ---  & 19.042 & 18.433 & 17.275 & 17.368 &     5.02$\pm$10.04 &    $-$8.77$\pm$10.04 &       \cr
03 28 16.47 & +48 29 41.9 &  ---  &  ---  & 18.986 & 18.067 & 17.075 & 16.986 &     9.54$\pm$6.66 &   $-$10.33$\pm$6.66 & detected in Z+Y \cr
03 30 17.49 & +48 04 56.8 &  ---  &  ---  & 19.040 & 18.184 & 17.283 & 17.416 &    13.82$\pm$7.76 &   $-$30.29$\pm$7.76 &       \cr
03 30 46.17 & +45 57 36.0 &  ---  &  ---  & 18.938 & 17.983 & 17.221 & 17.139 &     9.57$\pm$8.37 &    $-$9.08$\pm$8.37 & detected in Z+Y \cr
03 30 55.78 & +45 55 56.6 &  ---  &  ---  & 18.465 & 17.740 & 16.968 & 16.958 &    25.29$\pm$8.24 &   $-$19.76$\pm$8.24 & detected in Z+Y \cr
03 31 01.32 & +46 09 14.4 &  ---  &  ---  & 19.092 & 17.954 & 17.266 & 17.256 &     2.25$\pm$9.52 &    $-$0.70$\pm$9.52 & detected in Z+Y \cr
03 31 08.17 & +50 10 16.6 &  ---  &  ---  & 18.924 & 18.075 & 17.172 & 17.177 &    $-$2.82$\pm$9.25 &    $-$5.09$\pm$9.25 &       \cr
03 34 53.62 & +47 34 24.5 &  ---  &  ---  & 19.077 & 18.227 & 17.341 & 17.419 &    20.62$\pm$9.37 &    $-$4.52$\pm$9.37 & detected in Z+Y \cr
03 36 11.61 & +46 48 35.0 &  ---  &  ---  & 18.925 & 17.920 & 16.965 & 16.953 &    14.27$\pm$7.87 &   $-$21.11$\pm$7.87 & detected in Z+Y \cr
03 36 26.40 & +48 38 22.4 &  ---  &  ---  & 19.081 & 18.295 & 17.183 & 17.317 &    $-$2.90$\pm$9.79 &     0.75$\pm$9.79 & detected in Y \cr
03 43 15.74 & +47 34 45.0 &  ---  &  ---  & 18.859 & 17.886 & 17.177 & 17.267 &    22.76$\pm$8.90 &   $-$10.89$\pm$8.90 & detected in Z+Y \cr
 \hline
\end{tabular}
\end{table*}

%
%
\section{Table of substellar multiple system candidates in \APer{}}

\begin{table*}
  \caption{Coordinates (J2000), near-infrared ($ZYJHK$1$K$2) photometry, 
and proper motions (in mas/yr) for substellar multiple system candidates 
identified photmetrically in \APer{}
}
  \label{tab_APer:binary_candidates}
  \begin{tabular}{@{\hspace{0mm}}c c c c c c c c c c@{\hspace{0mm}}}
  \hline
R.A.\ & Dec.\  &  $Z$  &  $Y$  &  $J$  &  $H$  & $K$1 & $K$2 & $\mu_{\alpha}cos\delta$ & $\mu_{\delta}$ \\
 \hline
03 07 36.61 & +48 19 38.7 & 17.013 & 16.253 & 15.493 & 14.909 & 14.496 & 14.490 &    18.12$\pm$2.74 &   $-$24.09$\pm$2.74 \cr
03 18 40.74 & +50 56 01.1 & 16.252 & 15.537 & 14.801 & 14.229 & 13.793 & 13.764 &    20.63$\pm$3.05 &   $-$22.96$\pm$3.05 \cr
03 20 29.92 & +47 56 42.8 & 16.833 & 16.064 & 15.301 & 14.714 & 14.283 & 14.265 &    24.79$\pm$2.27 &   $-$23.82$\pm$2.27 \cr
03 23 08.69 & +48 04 50.5 & 16.699 & 16.046 & 15.294 & 14.734 & 14.318 & 14.353 &    17.92$\pm$2.28 &   $-$27.75$\pm$2.28 \cr
03 25 25.86 & +47 54 42.4 & 17.892 & 16.752 & 15.841 & 15.170 & 14.645 & 14.628 &    20.05$\pm$2.32 &   $-$25.97$\pm$2.32 \cr
03 27 31.32 & +48 39 23.1 & 16.692 & 15.920 & 15.161 & 14.620 & 14.165 & 14.140 &    27.28$\pm$2.26 &   $-$27.68$\pm$2.26 \cr
03 28 00.87 & +51 41 52.8 & 17.226 & 16.584 & 15.848 & 14.940 & 14.592 & 14.623 &    14.22$\pm$2.94 &   $-$20.31$\pm$2.94 \cr
03 30 24.28 & +51 54 10.8 & 18.011 & 16.808 & 15.836 & 15.211 & 14.622 & 14.618 &    28.01$\pm$2.96 &   $-$32.46$\pm$2.96 \cr
03 31 14.07 & +46 47 54.8 & 16.850 & 16.157 & 15.444 & 14.849 & 14.441 & 14.465 &    26.05$\pm$2.94 &   $-$24.76$\pm$2.94 \cr
03 33 37.35 & +50 43 39.5 & 14.641 & 14.275 & 13.598 & 12.386 & 12.259 & 12.598 &    15.57$\pm$2.86 &   $-$19.85$\pm$2.86 \cr
03 34 59.87 & +48 37 53.7 & 16.586 & 15.877 & 15.141 & 14.572 & 14.129 & 14.154 &    25.53$\pm$2.98 &   $-$25.35$\pm$2.98 \cr
03 35 47.37 & +49 17 42.9 & 16.817 & 15.913 & 15.158 & 14.590 & 14.151 & 14.167 &    24.20$\pm$3.05 &   $-$22.97$\pm$3.05 \cr
03 39 39.68 & +49 55 27.3 & 19.573 & 18.169 & 16.991 & 16.334 & 15.715 & 15.661 &    26.05$\pm$3.43 &   $-$21.38$\pm$3.43 \cr
03 40 59.57 & +47 11 41.2 & 16.554 & 15.897 & 15.149 & 14.565 & 14.132 & 14.149 &    23.90$\pm$2.94 &   $-$24.26$\pm$2.94 \cr
 \hline
\end{tabular}
\end{table*}

\label{lastpage}
\end{document}